\documentclass[nofootinbib,preprintnumbers,prd,superscriptaddress,twocolumn]{revtex4}
\usepackage{amsmath}
\usepackage{amsfonts}
\usepackage[utf8]{inputenc}
\usepackage[dvips]{graphicx}
\usepackage{graphicx}
\usepackage{color}
\usepackage{braket}
\usepackage{dcolumn}
\usepackage{bm,url}
\usepackage[linktocpage]{hyperref}
\usepackage{subfigure}
\usepackage{amsfonts}
\usepackage{comment}

\DeclareMathOperator{\sech}{sech}
\usepackage[usenames,dvipsnames,svgnames]{xcolor}  
\usepackage{hyperref}   
\definecolor{oxfordblue}{rgb}{0.0, 0.13, 0.28}
\definecolor{burgundy}{rgb}{0.5, 0.0, 0.13}
\definecolor{darkolivegreen}{rgb}{0.33, 0.42, 0.18}
\definecolor{darkblue}{rgb}{0,0,0.5}
\definecolor{richcarmine}{rgb}{0.84, 0.0, 0.25}
\definecolor{darkblue}{rgb}{0,0,0.5}
\definecolor{bluer}{rgb}{0.00,0.50,0.75}{}
\hypersetup{colorlinks=true, citecolor=blue, linkcolor=richcarmine,
urlcolor =magenta, filecolor=black}
\begin{document}

\newcommand{\tcr}{\textcolor{red}}
\newcommand{\tcb}{\textcolor{blue}}
\newcommand{\tcc}{\textcolor{cyan}}

\newcommand\be{\begin{equation}}
\newcommand\ee{\end{equation}}
\newcommand\bea{\begin{eqnarray}}
\newcommand\eea{\end{eqnarray}}
\newcommand\bseq{\begin{subequations}} 
\newcommand\eseq{\end{subequations}}
\newcommand\bcas{\begin{cases}}
\newcommand\ecas{\end{cases}}
\newcommand{\p}{\partial}
\newcommand{\f}{\frac}
\newcommand{\fixme}[1]{{\color{red}{\bf{#1}}}}
\newcommand{\fixmetoo}[1]{{\color{blue}{\bf{#1}}}}

\title{Physical Properties of Black Hole Solution in Einstein-Bel-Robinson Gravity}

\author{Seyed Naseh Sajadi}
\email{naseh.sajadi@gmail.com}
\affiliation{Strong Gravity Group, Department of Physics, Faculty of Science, Silpakorn University,\\ Nakhon Pathom 73000, Thailand}

\author{Supakchai Ponglertsakul}
\email{supakchai.p@gmail.com}
\affiliation{Strong Gravity Group, Department of Physics, Faculty of Science, Silpakorn University,\\ Nakhon Pathom 73000, Thailand}

\author{Dhruba Jyoti Gogoi}
\email{moloydhruba@yahoo.in}
\affiliation{Department of Physics, Moran College, Moranhat, Charaideo 785670, Assam, India}
\affiliation{Research Center of Astrophysics and Cosmology, Khazar University, Baku, AZ1096, 41 Mehseti Street, Azerbaijan}
\affiliation{Theoretical Physics Division, Centre for Atmospheric Studies, Dibrugarh University, Dibrugarh
786004, Assam, India}

\begin{abstract}
In this paper, we study the different properties of static spherically symmetric black hole solutions of Einstein-Bel-Robinson gravity (EBR), a modified four-dimensional theory of gravity quartic in curvature. We look at the orbit of massless and massive test bodies near a black hole, specifically computing the innermost stable circular orbit and photon sphere and finding them smaller than their Einstein counterparts in general relativity. Next, we obtain the deflection angle and shadow by an EBR black hole and find that both decreased compared to a non-rotating black hole in general relativity. We obtain a bound value for the coupling constant using the Shapiro time delay. Then, we explore the EBR black hole's lifetime and find that it decreases to Einstein's gravity. Quasinormal modes (QNMs) are computed using Pad\'e averaged 6th order WKB method showing that increasing the coupling constant lowers the damping rate of ring-down gravitational waves (GWs). The oscillation frequency of scalar QNMs decreases with the coupling constant, whereas it increases for electromagnetic QNMs. We also provide analytical rigorous bound of greybody factor. We show that the coupling constant has a small effect on the greybody factor. Finally, correspondence between greybody factor and quasinormal modes is also considered.
\end{abstract}

\maketitle

\section{Introduction}
Black holes are regions in spacetime where gravity is strong so that nothing can escape from them, not even light. Investigating black holes helps us to understand our universe and learn more about the structure of spacetime, also black holes are believed  to play an essential part in galaxy formations. General relativity (GR) proposed by Einstein is the main theory describing black hole physics. In GR, there are several distinct black hole solutions in four dimensions, for instance, static (Schwarzschild solution), static charged (Reissner-Nordstr\"om solution), rotating (Kerr solution), and charged rotating (Kerr-Newman solution) black holes.  The recent observations of the black hole shadows of M87 and SgrA by the EHT collaboration \cite{EventHorizonTelescope:2019dse, EventHorizonTelescope:2022wkp}, and the detection of the gravitational waves by the LIGO-VIRGO collaboration \cite{LIGOScientific:2016aoc}, confirm the existence of rotating black hole and predictions of Einstein's theory of gravity. 

However, despite its success, GR has serious problems on different scales. 
One of the ways to solve these problems is to modify Einstein's theory of gravity by adding higher-order terms in curvature. In recent years, various theories have been introduced and investigated \cite{Brans:1961sx, Sotiriou:2008rp, Capozziello:2011et, deRham:2014zqa}. One recently introduced theory is Einstein-Bel-Robinson gravity, which is quartic curvature theory \cite{Ketov:2022lhx}. This four-dimensional gravity is inspired by $M$-theory, which, at lower energies, is described by eleven-dimensional super-gravity. More specifically, there are suggestions that M-theory inspires this modified gravity compactified on a two-sphere factor represented by $S^3 \times S^4$ \cite{Ketov:2022lhx}.

Any modifications or alternates to GR have to be tested.
Testing the gravitational field near a compact massive object plays a crucial role in probing of theories of gravity \cite{Berti:2015itd, Allemandi:2006bm}. The motion of test particles in the closed vicinity of the event horizon of a black hole provides insight into the gravitational field and has contributed to constraining the spacetime parameters of theories. Particularly, the time-like circular geodesics around black holes are of great importance in astrophysics. These circular geodesics, like the innermost stable circular orbits (ISCO), are used to describe the accretion phenomenon around black holes. The light-like geodesics can also be used to show the effect of the strong gravitational field at small radii. The bending of light by gravity means that non-radial light rays can be captured by a black hole. The deflection of light and production of a lens-like effect are the results of gravitational lensing, which is caused by the curvature of spacetime around massive compact objects \cite{Virbhadra:1999nm}.

In \cite{Sajadi:2023bwe}, asymptotically anti-de Sitter (AdS)-black hole solutions to EBR gravity and their thermodynamic properties are investigated. Friedmann-Lemaitre-Robertson-Walker (FLRW) type solutions to the theory have also been considered \cite{Ketov:2022zhp}. Several asymptotically flat black hole solutions to the field equations have been explored \cite{CamposDelgado:2022sgc}. These solutions have been used to study phenomenological corrections to general relativistic predictions including light defection and shadows, slowly rotating black holes, quasinormal modes, and shadows with nonzero cosmological constant \cite{Belhaj:2023dsn, Arora:2023ijd, Davlataliev:2023ckw, Belhaj:2023pap, Hamil:2023neq}. The main goal of this paper is to investigate the effect of the parameter of theory on the null and time-dependent geodesics of the static black hole solution of the EBR gravity. Then, we study the stability of the black hole using quasi-normal modes.

The paper is organized as follows: In section \ref{bhsol}, we obtain the black hole solution for the generic value of the coupling constant of theory using the continued fraction expansion. We find the black hole solution for a small value of coupling constant and then study the time-like and null geodesics of a black hole. We derive an analytical expression for the deflection angle, shadow, and Shapiro time delay up to the first order of coupling constant $\beta$.  In section \ref{dynsta}, we investigate the dynamical stability of black holes using QNMs for scalar and electromagnetic perturbations. We also explore the BH's evolution via the evaporation process. 
In section \ref{grey}, we provide an exact rigourous bound of the greybody factor. In addition, relationship between greybody factor and QNMs is also investigated. Finally, in section \ref{concl}, we present our conclusions.

\section{Black hole solution}\label{bhsol}
In four dimensions, Einstein-Bel-Robinson gravity  is determined by the following action \cite{Ketov:2022lhx}
\begin{equation}\label{eq1}
\mathcal{S}=\dfrac{1}{16\pi G}\int d^{4}x\sqrt{-g}\left[R-2\Lambda-\beta \left(\mathcal{P}^2-\mathcal{G}^2\right)\right],
\end{equation}
where $R$ is the Ricci scalar, $\beta$ is the coupling constant of the theory, and $\Lambda=-3/\ell^2$, where $\ell$ refers to the curvature radius of the maximally symmetric
anti-de Sitter (AdS) solution of the field equations. The quantities $\mathcal{G}$ and $\mathcal{P}$
\begin{align}
\mathcal{P}=&\dfrac{1}{2}\sqrt{-g}\epsilon_{\mu \nu \rho \sigma}R^{\rho \sigma}{}_{\alpha \beta}R^{\mu \nu \alpha \beta},\nonumber\\
\mathcal{G}=&R^{2}-4R_{\mu\nu}R^{\mu\nu}+R_{\mu\nu\rho\sigma}R^{\mu\nu\rho\sigma},
\end{align}
are the Euler and Pontryagin topological densities respectively. These come from the square of the Bel-Robinson tensor \cite{Ketov:2022lhx, Sajadi:2023bwe}.

Varying the action \eqref{eq1} with respect to the metric tensor yields the following equations of motion
\begin{align}\label{fieldeq3}
\mathcal{E}_{a b}&\equiv R_{a b}-\dfrac{1}{2}g_{a b}R+\Lambda g_{a b} -\beta \mathcal{K}_{a b} = 0,
\end{align}
where 
\begin{equation}\label{Kab}
\mathcal{K}_{a b}=\mathcal{K}^{\mathcal{G}}_{a b}+\mathcal{K}^{\mathcal{P}}_{a b},
\end{equation}
with the Pontryagin topological term 
\begin{equation}
\begin{gathered}\label{KabG}
	\mathcal{K}^{\mathcal{G}}_{a b}=\frac{1}{2} g_{ab} \mathcal{G}^2-2\Big[2 \mathcal{G} R R_{ab}-4 \mathcal{G} R_{a}^\rho R_{b \rho}
	+2 \mathcal{G} R_a^{\rho \sigma \lambda} R_{b \rho \sigma \lambda}\\
 +4 \mathcal{G} R^{\rho \sigma} R_{a \rho \sigma b}+2 g_{ab} R \square \mathcal{G}-
2 R \nabla_a \nabla_b \mathcal{G}
	-4 R_{ab} \square \mathcal{G}+\\4\left(R_{a \rho} \nabla^\rho \nabla_b \mathcal{G}+R_{b \rho} \nabla^\rho \nabla_a \mathcal{G}\right) -4 g_{ab} R_{\rho \sigma} \nabla^\sigma \nabla^\rho \mathcal{G}+\\4 R_{a \rho b \sigma} \nabla^\sigma \nabla^\rho \mathcal{G}\Big],
	\end{gathered}
\end{equation}
and the Euler term
\begin{equation}
\begin{gathered}\label{KabP}
\mathcal{K}^{\mathcal{P}}_{a b}=-\dfrac{1}{2}g_{ab}\mathcal{P}^2+\epsilon_{c d e f}g_{ab}\mathcal{P}R_{\alpha\beta}{}^{ef}R^{\alpha\beta cd}\\-2\mathcal{P}\epsilon_{b\alpha de}R_{\beta c}{}^{de}R_{a}{}^{\alpha \beta c}-2\mathcal{P}\epsilon_{a \alpha de}R_{\beta c}{}^{de}R_{b}{}^{\alpha \beta c}
		\\- 2\mathcal{P}\epsilon_{b\beta c d}\nabla_{\alpha}\nabla^{d}R_{a}{}^{\alpha\beta c}-2\mathcal{P}\epsilon_{a\beta cd}\nabla_{\alpha}\nabla^{d}R_{b}{}^{\alpha\beta c}\\-2\epsilon_{b\alpha cd}\nabla_{\beta}R_{a}{}^{\beta cd}\nabla^{\alpha}\mathcal{P}-2\epsilon_{a\alpha cd}\nabla_{\beta}R_{b}{}^{\beta cd}\nabla^{\alpha}\mathcal{P}
		 \\-2\epsilon_{b\alpha cd}R_{a\beta}{}^{cd}\nabla^{\beta}\nabla^{\alpha}\mathcal{P}-2\epsilon_{a\alpha cd}R_{b\beta}{}^{cd}\nabla^{\beta}\nabla^{\alpha}\mathcal{P}\\-2\epsilon_{b\beta cd}\nabla^{d}R_{a\alpha}{}^{\beta c}\nabla^{\alpha}\mathcal{P}
	2\epsilon_{a\beta cd}\nabla^{d}R_{b\alpha}{}^{\beta c}\nabla^{\alpha}\mathcal{P}\,.
\end{gathered}
\end{equation}
In spherically symmetric spacetime, $\mathcal{P}$ has zero contribution to the field equation \cite{Sajadi:2023bwe}, therefore we shall not consider the effect of $\mathcal{P}$ any further. 

The most general radially symmetric metric can be written in the form \cite{Sajadi:2023bwe}
\begin{equation}\label{sssmet}
ds^2 = -N(r) f(r) dt^2 + \frac{dr^2}{f(r)} + r^2 \left(d\theta^2 + \frac{\sin^2(\sqrt{k}\theta)}{k} d\phi^2\right),
\end{equation}
where $k\in \lbrace 1,0,-1\rbrace$ correspond to spherical, planar, and hyperbolic transverse sections. The metric functions $N(r)$ and $f(r)$ must be determined from the field equations  \eqref{fieldeq3}. In the next section, we obtain the black hole solution for the theory using continued fraction expansion \cite{Kokkotas:2017zwt} and the thermodynamics of black holes. 

\subsection{Exact Black Hole Solution}
\begin{figure*}
	\centering
	\includegraphics[width=0.6\columnwidth]{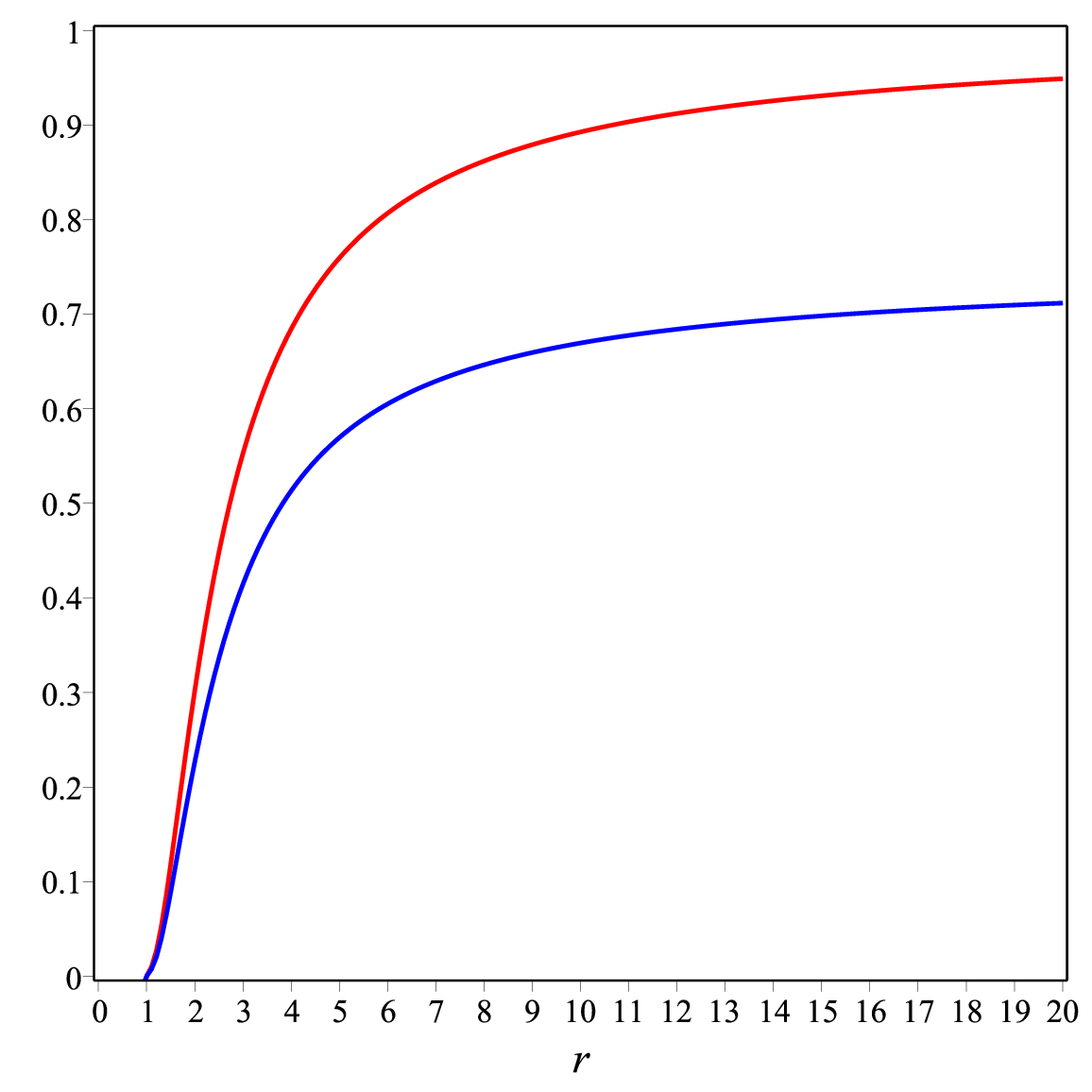}
	\includegraphics[width=0.6\columnwidth]{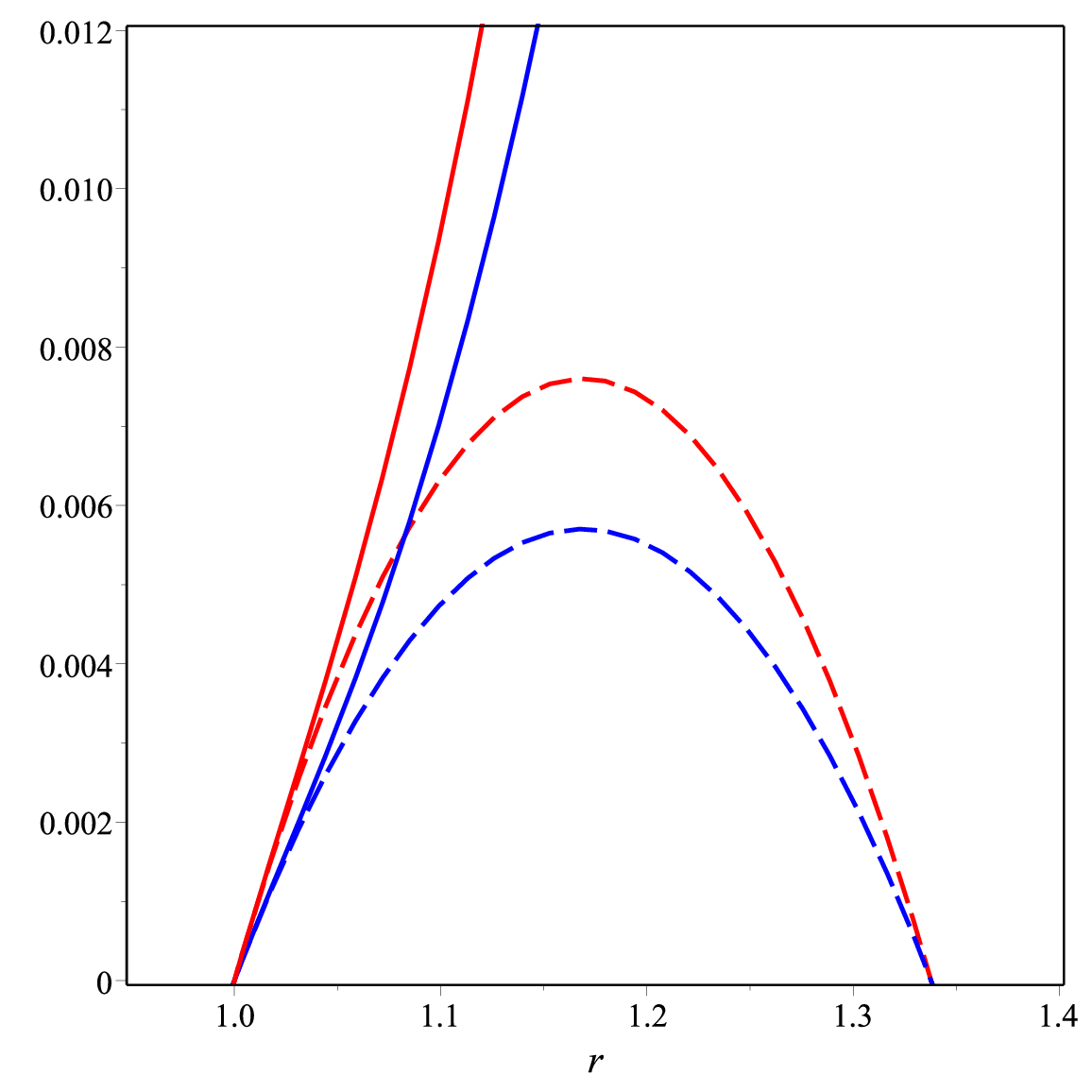}
    \includegraphics[width=0.6\columnwidth]{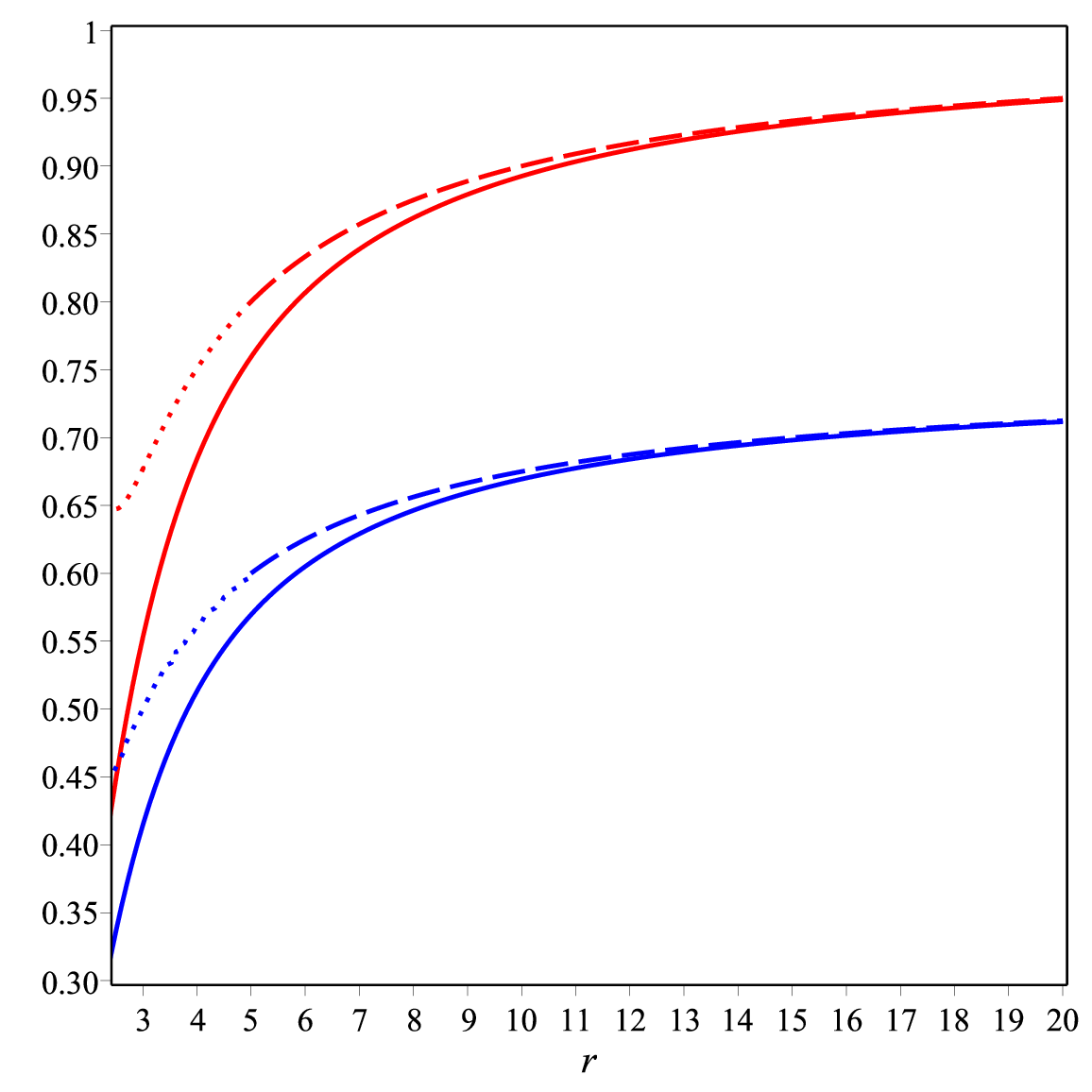}
	\caption{The behavior of $f(r)$ (\textcolor{red}{solid red line}) and $0.75h(r)$ (\textcolor{blue}{blue solid line}) in terms of $r$ for $r_{+}=1, \beta=0.5$. The left figure is the full continued fraction expansion solution, the middle figure compares this solution to the near horizon approximation and the right figure compares this solution to the large-$r$ approximation.}
	\label{frrpl}
\end{figure*}

Here, we want to obtain the black hole solution for the EBR theory with a generic coupling constant $\beta$. To do so, we expand the metric functions in the large $r$ as follows:
\begin{align}\label{eqqqhfo}
N(r)=&\sum_{n=0}\dfrac{\mathcal{N}_{n}}{r^{n}}=\mathcal{N}_{0}+\dfrac{448\mathcal{N}_{0}\beta\mathcal{F}_{1}^{3}}{r^{9}}+\mathcal{O}(r^{-11}),\\
 f(r)=&\sum_{n=0}\dfrac{\mathcal{F}_{n}}{r^{n}}=1+\dfrac{\mathcal{F}_{1}}{r}-\dfrac{576\beta \mathcal{F}_{1}^{3}}{r^{9}}-\dfrac{536\beta \mathcal{F}_{1}^{4}}{r^{10}} \nonumber \\
 &+\mathcal{O}(r^{-11}).
\end{align}
Here, $\mathcal{N}_i$ and $\mathcal{F}_i$ are determined from the field equations \eqref{Kab}.
Expanding the function $f(r)$ and $h(r)\equiv N(r)f(r)$ around the event horizon ($r_+$), one gets
\begin{align}\label{eq7}
h(r) &= h_{1}(r-r_{+})+h_{2}(r-r_{+})^{2}+h_{3}(r-r_{+})^{3}+\cdots,\\
f(r)  &= f_{1}(r-r_{+})+f_{2}(r-r_{+})^{2}+f_{3}(r-r_{+})^{3}+\cdots
\label{eq8}
\end{align}
Then inserting these expressions into \eqref{fieldeq3}, we find
\begin{widetext}
\begin{align}\label{eq9}
f_{2}&=\dfrac{1}{16\beta h_{1}(1+f_{1}r_{+})}\Big[32f_{1}\beta r_{+}f_{1}^3-48\beta h_{2}f_{1}-h_{1}r_{+}^4-16\beta f_{1}^2(3r_{+}h_{2}-2h_{1})\pm h_{1}r_{+}^2A\Big],\\
h_{2}&=-\dfrac{h_{1}}{128r_{+}\beta f_{1}^2(Af_{1}r_{+}^4+Ar_{+}^3-f_{1}r_{+}^6-r_{+}^5-96\beta f_{1}^4r_{+}^3-384\beta f_{1}^3r_{+}^2-480\beta r_{+}f_{1}^2-192\beta f_{1})}\times\nonumber\\
&\Big(r_{+}^7A-9f_{1}r_{+}^{10}+9Af_{1}r_{+}^{8}+32A\beta f_{1}^4r_{+}^5+640A\beta f_{1}^3r_{+}^4+544A\beta f_{1}^2r_{+}^3-64A\beta r_{+}-128Af_{1}\beta r_{+}^2\nonumber\\
&-448\beta f_{1}^2r_{+}^5-6144f_{1}^3\beta^2 +896\beta f_{1}^4r_{+}^7+6144\beta^2 f_{1}^7 r_{+}^4+448\beta r_{+}^6f_{1}^3+30720\beta^2 r_{+}^3f_{1}^6+36864\beta^2 \nonumber\\
&r_{+}^2f_{1}^5+6144\beta^2 f_{1}^4r_{+}-r_{+}^9\Big),
\end{align}
\end{widetext}
where $\small{A=\sqrt{r_{+}^4-192\beta f_{1}^2(f_{1}r_{+}+1)}}$. The $f_{1}$ and $h_{1}$ are undetermined integration constants. We want to obtain an approximate analytic solution which valid near the horizon and at large $r$. 
To this end, we employ a continued fraction expansion and write \cite{Rezzolla:2014mua, Sajadi:2025nkm, Sajadi:2020axg}
\begin{equation}\label{eq17}
f(r)=\dfrac{xA(x)}{N(r)},\hspace{0.5cm}N(r)=B^{2}(x),\hspace{0.5cm} x= 1- \frac{r_+}{r},
\end{equation}
with
\begin{align}
A(x) &=1-\epsilon(1-x)+(a_{0}-\epsilon)(1-x)^{2}\nonumber\\
&~~~~+\tilde{A}(x)(1-x)^{3},\\
\label{Ax}
B(x) &=1+b_{0}(1-x)+\tilde{B}(x)(1-x)^{2},
\end{align}
where
\begin{equation}
\tilde{A}(x)=\dfrac{a_{1}}{1+\dfrac{a_{2}x}{1+\dfrac{a_{3}x}{1+\dfrac{a_{4}x}{1+...}}}},\;\; 
 \tilde{B}(x)=\dfrac{b_{1}}{1+\dfrac{b_{2}x}{1+\dfrac{b_{3}x}{1+\dfrac{b_{4}x}{1+...}}}}.
\label{cfrac}
\end{equation}
We truncate the continued fraction at an order of $4$.
By expanding (\ref{eq17}) near the horizon ($ x\to 0 $) and
the asymptotic  region ($ x\to 1 $), we obtain  
\begin{equation}
\epsilon=-\dfrac{\mathcal{N}_{0}\mathcal{F}_{1}}{r_{+}}-1,  \qquad b_{0}=\dfrac{\mathcal{F}_{1}-\mathcal{N}_{0}\mathcal{F}_{1}}{2r_{+}},\;\;a_{0}=0,
\end{equation}
for the lowest-order expansion coefficients, with the remaining
$a_i$ and $b_i$ given in terms of $(r_+, h_1, f_1)$; we provide these expressions explicitly in the Appendix \ref{sec:Appendix}. 
For a static space-time, the temperature of a black hole is given by
\begin{equation}
    T=\left.\dfrac{1}{4\pi}\dfrac{(Nf)^{\prime}}{\sqrt{N}}\right\vert_{r_{+}}=\dfrac{\sqrt{f_{1}h_{1}}}{4\pi},
\end{equation}
where prime denotes derivative with respect to $r$. The entropy can be computed as follows
\begin{align}\label{eqqEnt}
S&=-2\pi\int_{Horizon}d^{2}x\sqrt{\eta}\dfrac{\delta L}{\delta R_{a b c d}}\epsilon_{a b}\epsilon_{c d},\nonumber\\
&=
\left(\pi r_{+}^2\left[1- {\dfrac{32\beta}{r^{4}}\left(f^{\prime\prime}-f^{\prime 2}+\dfrac{3f^{\prime}N^{\prime}}{2N}\right)}\right] \right)_{r_+},\nonumber\\
&=\pi r_{+}^2\left[1-\beta\left[\dfrac{16f_{2}}{r_{+}^{4}}+\dfrac{48f_{1}h_{2}}{h_{1}r_{+}^4}-\dfrac{32f_{1}^2}{r_{+}^{4}}\right]\right].
\end{align}
We can also regard the coupling $\beta$ as a thermodynamic parameter, whose
conjugate is denoted as $\psi_{\beta}$. We find
\begin{equation}
    \psi_{\beta}=-4\pi\left.\dfrac{h^{\prime}f(f-1)}{\sqrt{hf}}\right\vert_{r_{+}}=-4\pi\sqrt{f_{1}h_{1}}.
\end{equation}
From the Smarr formula, we can obtain the mass as follows
\begin{align}
    M&=2TS+6{\psi_{\beta} \beta}, \nonumber \\
    &=\dfrac{\sqrt{f_{1}h_{1}}r_{+}^2}{2}\left[1-\beta\left(\dfrac{16f_{2}}{r_{+}^{4}}+\dfrac{48f_{1}h_{2}}{h_{1}r_{+}^4}-\dfrac{32f_{1}^2}{r_{+}^{4}}\right)\right] \nonumber \\
    &~~~~-2\beta \sqrt{f_{1}h_{1}},
\end{align}
yielding the mass parameter as a function of the horizon radius and the coupling constant.
From the first law of thermodynamics of black holes, we get
\begin{equation}
    dM=TdS+\psi_{\beta}d\beta.
\end{equation}
We now impose the first law, which becomes
\begin{equation}
    \dfrac{\partial M}{\partial r_{+}}dM+\dfrac{\partial M}{\partial \beta}d\beta=T\dfrac{\partial S}{\partial r_{+}}dr_{+}+T\dfrac{\partial S}{\partial\beta}d\beta+\psi_{\beta}d\beta.
\end{equation}
We assume $h_{1}=f_{1}$ to solve the differential equations. Therefore, we get
\begin{align}
    f_{1}(r_{+},\beta)&=-\dfrac{1}{3r_{+}}+\dfrac{2c_{1}\beta^{\frac{1}{3}}}{3r_{+}\mathcal{A}^{\frac{1}{3}}}+\dfrac{\mathcal{A}^{\frac{1}{3}}}{6r_{+}c_{1}\beta^{\frac{1}{3}}},\nonumber\\
     \mathcal{A}&=-8\beta c_{1}^{3}+108r_{+}^{6}+12r_{+}^{3}\sqrt{81r_{+}^6-12\beta c_{1}^{3}},
    \end{align}
where $c_1$ is integration constant. Now, we insert this expression of $f_1$ into \eqref{eq17}, one can obtain the full expression of the metric functions. In Fig.~\ref{frrpl}, the two metric functions are shown where the event horizon is located at $r_+=1$. It should be noted that since we assume $f_{1}=h_{1}$, therefore $b_{1}=0$ which leads to $f(r)=h(r)$. 

\subsection{Approximated BH solution}
In the previous section, we obtain a static exact solution for generic $\beta$. This section attempts to obtain a perturbative solution for small $\beta$. To do this, we consider expansions of the form
\begin{align}\label{betapprox}
N(r) &= N_0 (r) + \beta N_1(r) + \beta^2 N_2(r) + \cdots,\nonumber\\ f(r) &= f_0 (r) + \beta f_1(r) + \beta^2 f_2(r) + \cdots
\end{align}
With these expansions, we solve \eqref{fieldeq3} perturbatively in $\beta$. Solving the $tt$ and $rr$ components of  \eqref{fieldeq3} yields \cite{Sajadi:2023bwe}
\begin{align}
N(r) &=  1 -\dfrac{3584 m^{3} \beta}{r^{9}}+\Big(\frac{4236115968  k m^{5}}{17 r^{17}}
\nonumber\\
&~~~~-\frac{569393152 m^{6}}{r^{18}}
\Big) \beta^2,
\label{Nbetsol} \\
f(r) & =k-\frac{2 m}{r} +\Big(\frac{4608 k m ^{3}}{r^{9}}-\frac{8576 {m}^{4}}{r^{10}}\Big)\beta\nonumber\\
&~~~~-\Big(\frac{283115520 k^{2} m^{5}}{r^{17}}-\frac{20514373632 k m^{6}}{17 r^{18}}\nonumber\\
&~~~~+\frac{1275707392 m^{7}}{r^{19}}
\Big)\beta^2,
\label{fbetsol} 
\end{align}
where $m$ is a constant of integration.  It is straightforward to check that the remaining field equations are also satisfied. We see that
$N$ is not constant even in linear order in $\beta$. Unlike the generic $\beta$ case, a perturbative solution obtained in this section is found with $g_{tt}g_{rr}\neq -1$.
In the next section, we study the null and time-like geodesics to explore how they are modified compared to the Einstein gravity case.

\subsection{Particle motion around black hole}
In this section, we study the timelike and null geodesics of particle around the solutions \eqref{betapprox}. In particular, we derive the geodesics equation in the equatorial plane ($\theta=\pi/2$). Then, we compute the innermost
stable circular orbit (ISCO) for timelike geodesics. In addition, we explore the behavior of photon rings, deflection angle, and shadow for null geodesics.

\begin{figure}
	\centering
	\includegraphics[width=0.95\columnwidth]{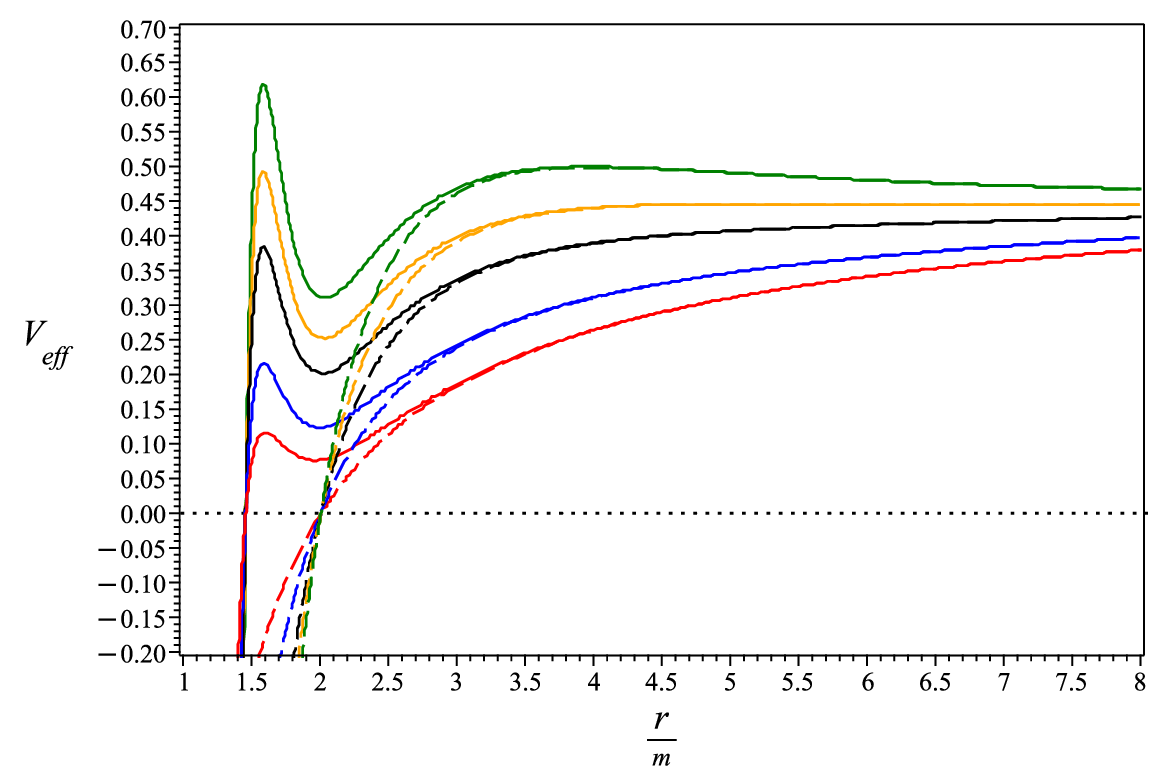}~~~~~~~~~~
	\caption{Plots of $V_{eff}$ in terms of $r/m$ for $L/m=\textcolor{red}{1},\textcolor{blue}{2},3,\textcolor{orange}{3.5},\textcolor{green}{4}$ and $ \beta/m^6=0.2$ (bottom to top). The dashed lines denote Einstein's gravity and the EBR gravity is solid lines.}
	\label{veff1}
\end{figure}

Since the spacetime is independent of $ t $ and $ \phi $, there are two conserved quantities:
\begin{equation}
E=h(r) \dot{t},\hspace{0.5cm}L=r^{2}\dot{\phi},
\end{equation}
{where $h(r)=N(r)f(r)$.}
Without loss of generality, we consider the geodesics equation
 on the equatorial plane ($ \theta=\frac{\pi}{2} $). The radial geodesics equation is  
\begin{equation}\label{geod}
\dfrac{1}{2}\dfrac{h}{f}\dot{r}^{2}+\dfrac{1}{2}\left[ h\left( \dfrac{L^{2}}{r^{2}}+\mu\right)\right]=\dfrac{E}{2}.
\end{equation}
Here $\mu=1$ is for a massive particle and $\mu=0$ is for a massless particle. The second term on the left-hand side can be regarded as the effective potential. Note that, the kinetic energy term ($\dot{r}^2$) has a non-canonical normalization factor.
Since we have (up to linear order in $\beta$ and $k=1$) \begin{equation}\label{eqapproximat}
h(r)=1-\dfrac{2m}{r}+\left(\dfrac{1024m^3}{r^9}-\dfrac{1408m^4}{r^{10}}\right)\beta.
\end{equation}
Thus, the effective potential for a massive particle is 
\begin{align}\label{Veff38}
V_{eff} &=\dfrac{1}{2}\Big[ h\Big( \dfrac{L^{2}}{r^{2}}+1\Big)\Big], \nonumber \\
&=\dfrac{1}{2}\left[1-\dfrac{2m}{r}+\Big(\dfrac{1024m^3}{r^9}-\dfrac{1408m^4}{r^{10}}\Big)\beta\right]\left( \dfrac{L^{2}}{r^{2}}+1\right).
\end{align} 
From \eqref{Veff38}, one can deduce the shape of the effective potential with the presence of the BR term ($\beta-$ dependence term). The BR part consists of positive and negative terms which are the repulsive and attractive potentials that dominate at small $r$. The positive term creates a minimum and the negative term creates a maximum in the effective potential curve. Therefore, one expects the BR term to make extra extremums in the effective potential with respect to Einstein's gravity. This behavior is illustrated in Fig.~\ref{veff1}. 

\begin{figure*}
	\centering
	\includegraphics[width=0.95\columnwidth]{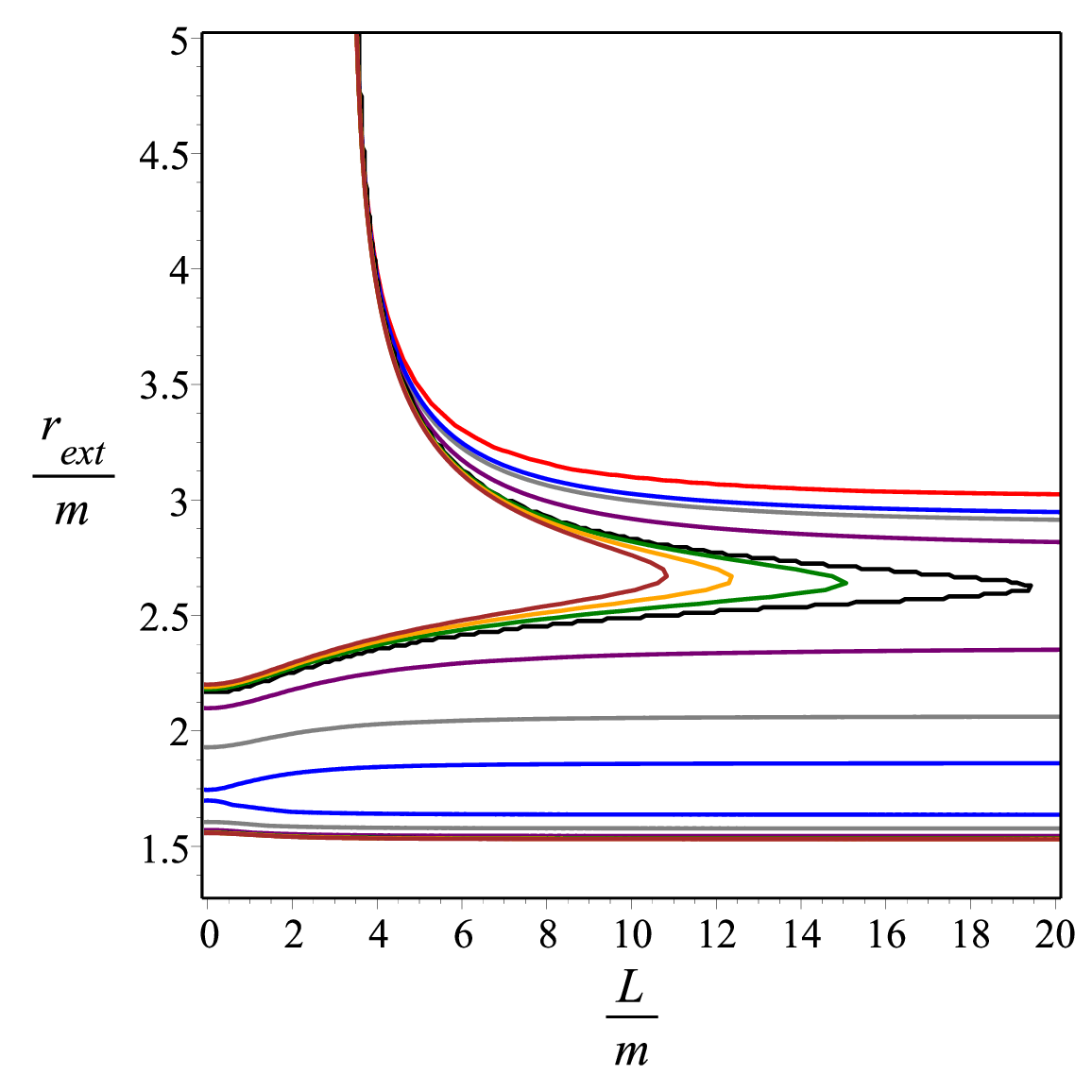}~~~~~~~~~~
	\includegraphics[width=0.95\columnwidth]{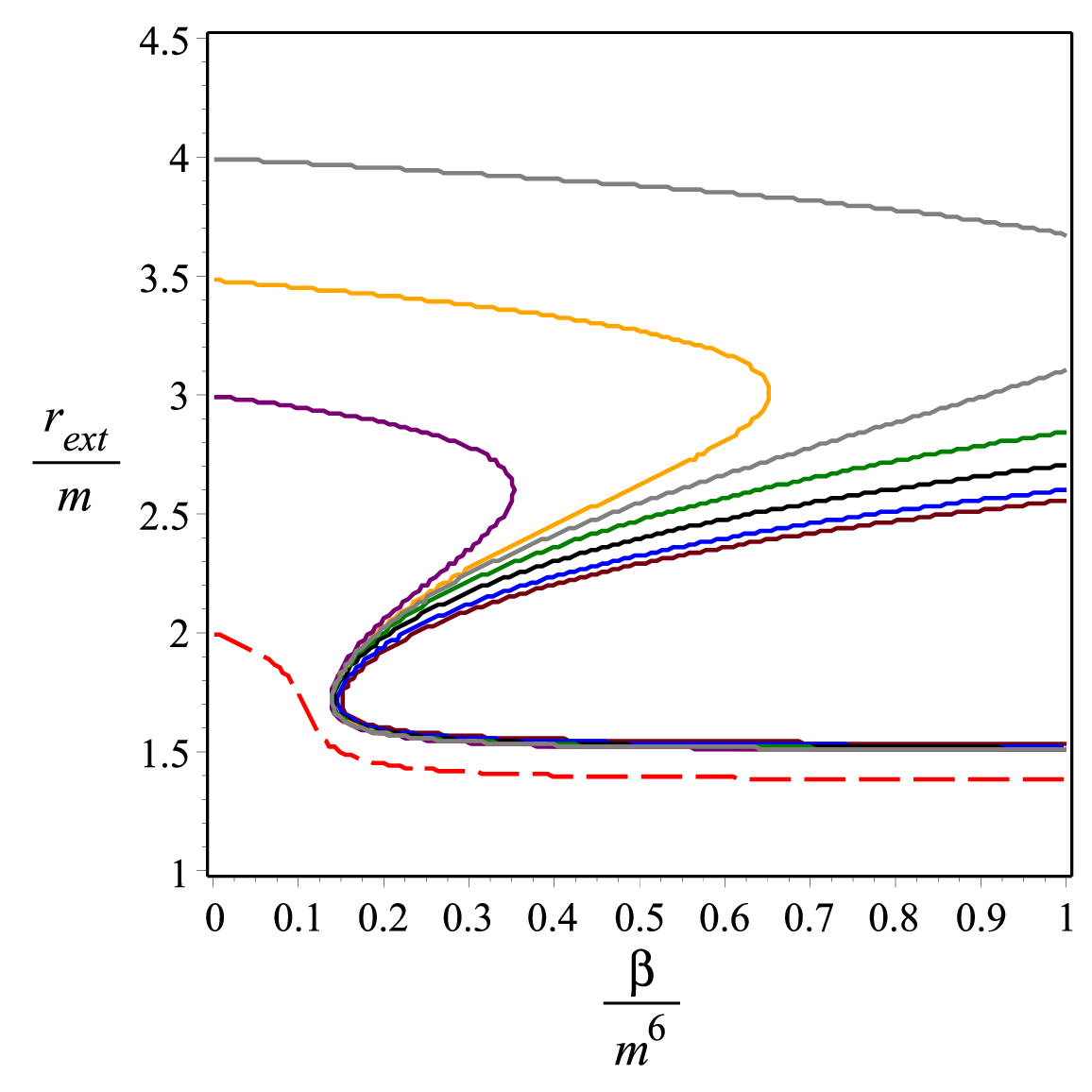}
	\caption{Left: The behavior of $\frac{r_{ext}}{m}$ in terms of $\frac{L}{m}$ for $\frac{\beta}{m^6}=\textcolor{red}{0},\textcolor{blue}{0.15},\textcolor{gray}{0.2},\textcolor{black}{0.3},0.362,\textcolor{green}{0.37},\textcolor{orange}{0.38},0.39$ for massive particles. Right: The behavior of $\frac{r_{ext}}{m}$ and $\frac{r_{+}}{m}$ (red dashed line) in terms of $\frac{\beta}{m^6}$ for $\frac{L}{m}=\textcolor{blue}{0},1,\textcolor{green}{2},3,4,5,100$.}
	\label{rextlb}
\end{figure*}

Now, we consider the case of circular, time-like geodesics. The conditions for circular geodesics are $V_{eff}=V_{eff}^{\prime}=0$.
The stability of the circular orbit is deduced from the sign of $V_{eff}^{\prime\prime}$, with a positive sign indicating stability and a negative sign indicating instability.
In Fig.~\ref{veff1}, we plot $V_{eff}$ for $\beta/m^6=0.2$ with different values of $L$. For a large value of $L$, there are two extrema in the curve of $V_{eff}$, with the maximum (minimum) rendering the unstable (stable) orbits. By decreasing the value of $L$, the radius of the unstable equilibrium orbit increases and the radius of the stable equilibrium orbit decreases. 

In Fig.~\ref{rextlb}, the behavior of extremes of the effective potential, i.e., $V'(r_{ext})=0,$ has been shown. For fixed $\beta/m^6$, the number of extrema increases with $L/m$. The ISCO is at the inflection point of $V_{eff}$, i.e., for which $V_{eff}^{\prime\prime}= 0$. It is possible to find approximation formula for $r_{isco},L_{isco}$ such that $V'_{eff}(r_{isco},L_{isco})=V''_{eff}(r_{isco},L_{isco})=0$.

We expand the following up to first order in $\beta$, $r_{isco}=r^{0}_{isco}+\beta r^{1}_{isco} $ and $L_{isco}=L^{0}_{isco}+\beta L^{1}_{isco}$. Substituting these into the equations $V_{eff}^{\prime}=V_{eff}^{\prime\prime}=0$ and collecting in powers of $\beta$, one obtains 
\begin{align}
    r_{isco} &= 6m-\dfrac{1871\beta}{26244 m^{5}}+\mathcal{O}(\beta^2),\\
    L_{isco} &= 2\sqrt{3}m -\dfrac{227\sqrt{3}\beta}{78732m^5}+\mathcal{O}(\beta^2).
\end{align}

\begin{figure}
	\centering
	\includegraphics[width=0.95\columnwidth]{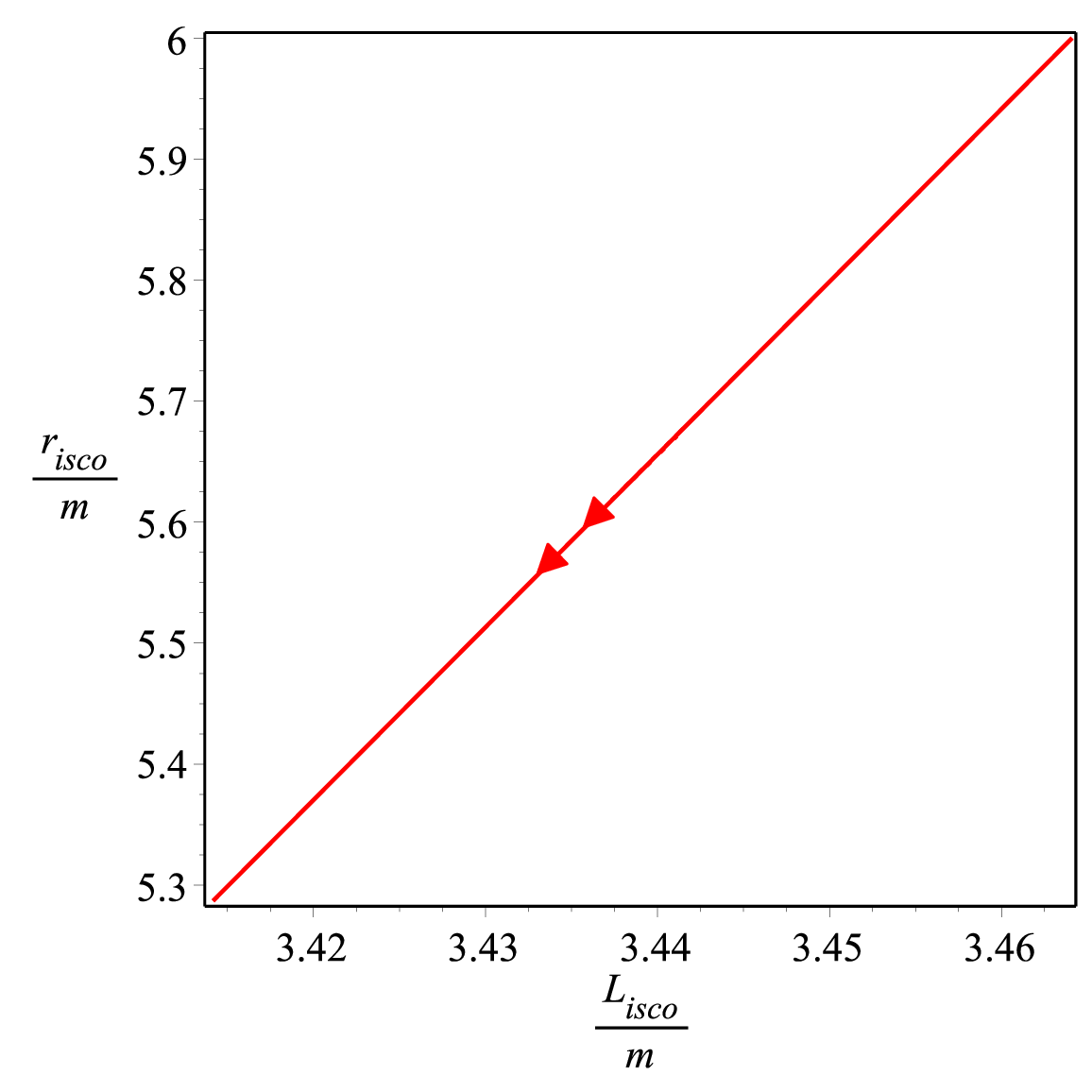}~~~~~~~~~~
	\caption{The behavior of $r_{isco}/m$ in terms of $L_{isco}/m$. Arrows show the direction of increasing $\beta/m^6$.}
	\label{rliscom}
\end{figure}
The corrections to the corresponding values in general relativity, i.e., $r_{isco} = 6m$ and $L_{isco}=2\sqrt{3}m$, have been obtained. In Fig.~\ref{rliscom}, the behavior of $r_{isco}/m$ in terms of $L_{isco}/m$ has been displayed. As can be seen, by increasing $\beta/m^6$ the ISCO quantities are relatively smaller than those of the Einstein gravity.  The arrows show the direction of increasing $\beta/m^6$.
The condition for a vanishing radial velocity i.e., $\dot{r}^2=0$ implies $E = V_{eff}$.
Hence, the total energy of a particle, in the case of ISCO orbits, reads
\begin{align}
        E_{isco}=\dfrac{8}{9}-\dfrac{95\beta}{177147m^{6}}+\mathcal{O}(\beta^2). \label{Eisco}
\end{align}
The coordinate period of the time-like orbits is given by \cite{raine2015black} 
\begin{align}
    T^{(\tau)}_{isco}=\left.\dfrac{2\pi E r^{2}}{L h(r)}\right\vert_{r_{isco}}=16\sqrt{3}\pi m -\dfrac{5386\sqrt{3}\pi \beta}{19683 m^{5}}
    +\mathcal{O}(\beta^2). \label{Tisco}
\end{align}
By setting $\beta=0$ in \eqref{Eisco} and \eqref{Tisco}, we recover ISCO quantities obtained in general relativity.

\begin{figure}
	\centering
	\includegraphics[width=0.95\columnwidth]{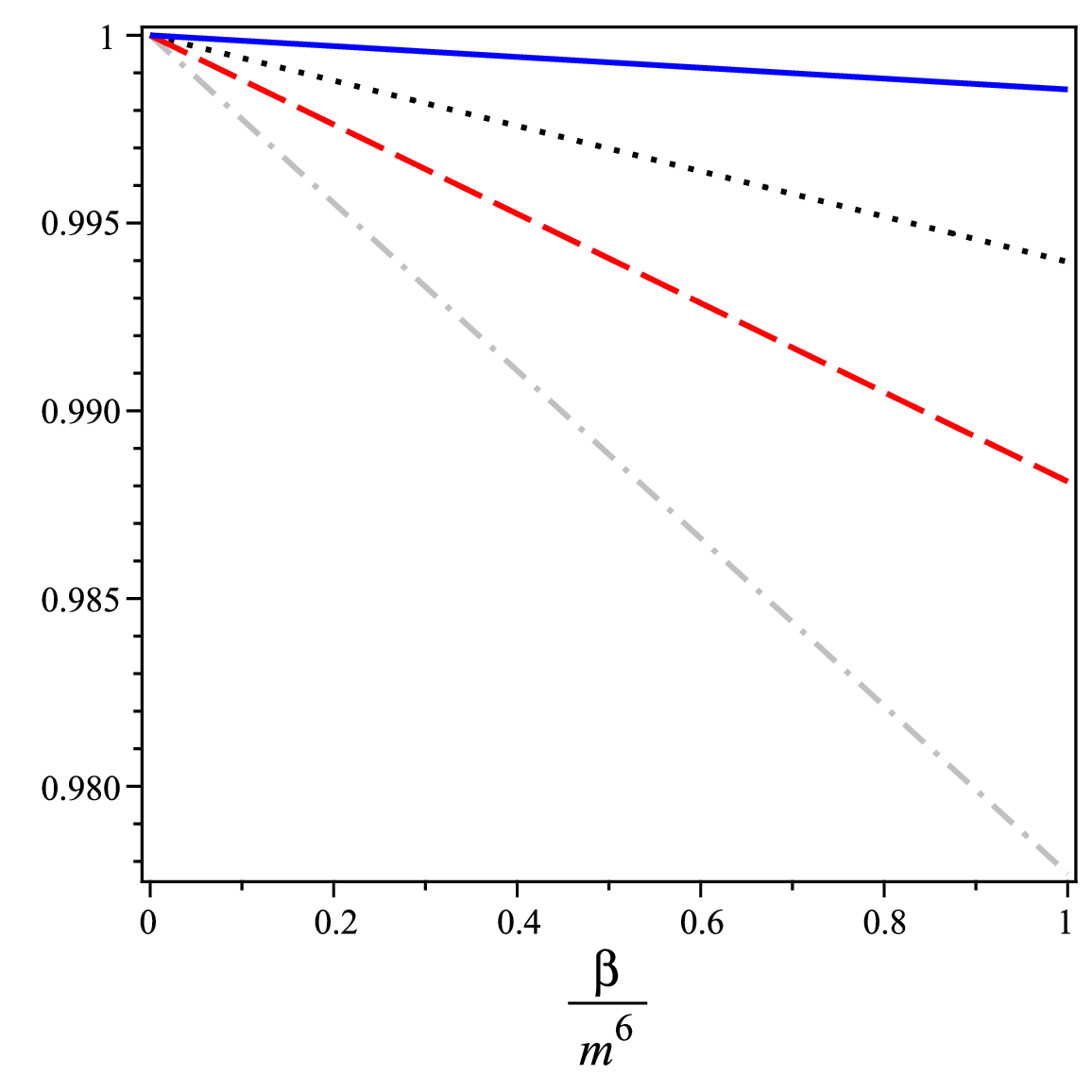}~~~~~~~~~~
	\caption{The behavior of ISCO quantities in terms of coupling of theory. The behavior of $L_{isco}/2\sqrt{3}m$ (solid blue line), $r_{isco}/6m$ (dashed red line), $9E_{isco}/4$ (dotted line) and $T_{isco}/12\sqrt{3}\pi m$ (gray dash-dotted line). }
	\label{veff00}
\end{figure}

The results for each ISCO quantity are shown in Fig.~\ref{veff00}. The corrections due to the BR term become the most significant for small masses, and we can see that by increasing $\beta$, ISCO quantities deviate more from GR.

\begin{figure}
	\centering
	\includegraphics[width=0.95\columnwidth]{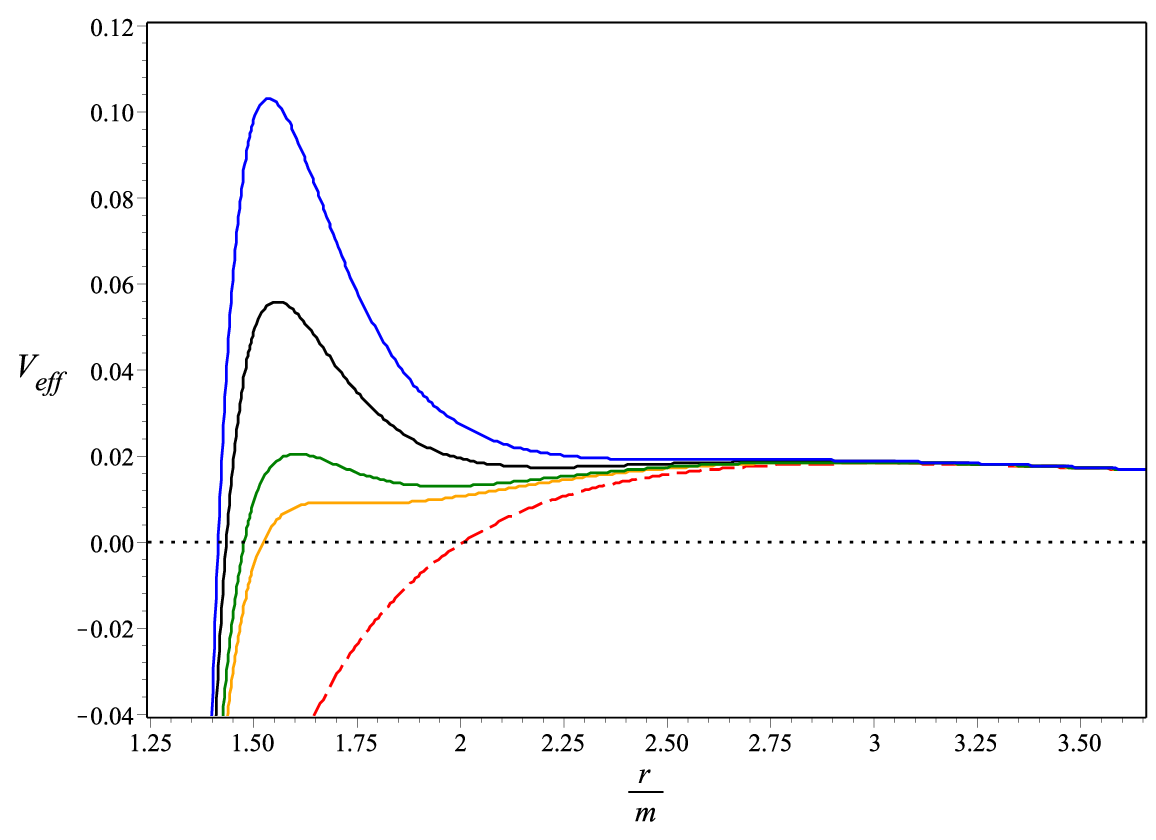}~~~~~~~~~~
	\caption{Plots of $V_{eff}$ in terms of $r/m$ for $ \beta/m^6=\textcolor{red}{0},\textcolor{orange}{0.14},\textcolor{green}{0.25},0.35,\textcolor{blue}{0.4}$ (bottom to top). The dashed lines for Einstein gravity and solid lines for EBR gravity.}
	\label{veffnull}
\end{figure}

\begin{figure*}
	\centering
	\includegraphics[width=0.95\columnwidth]{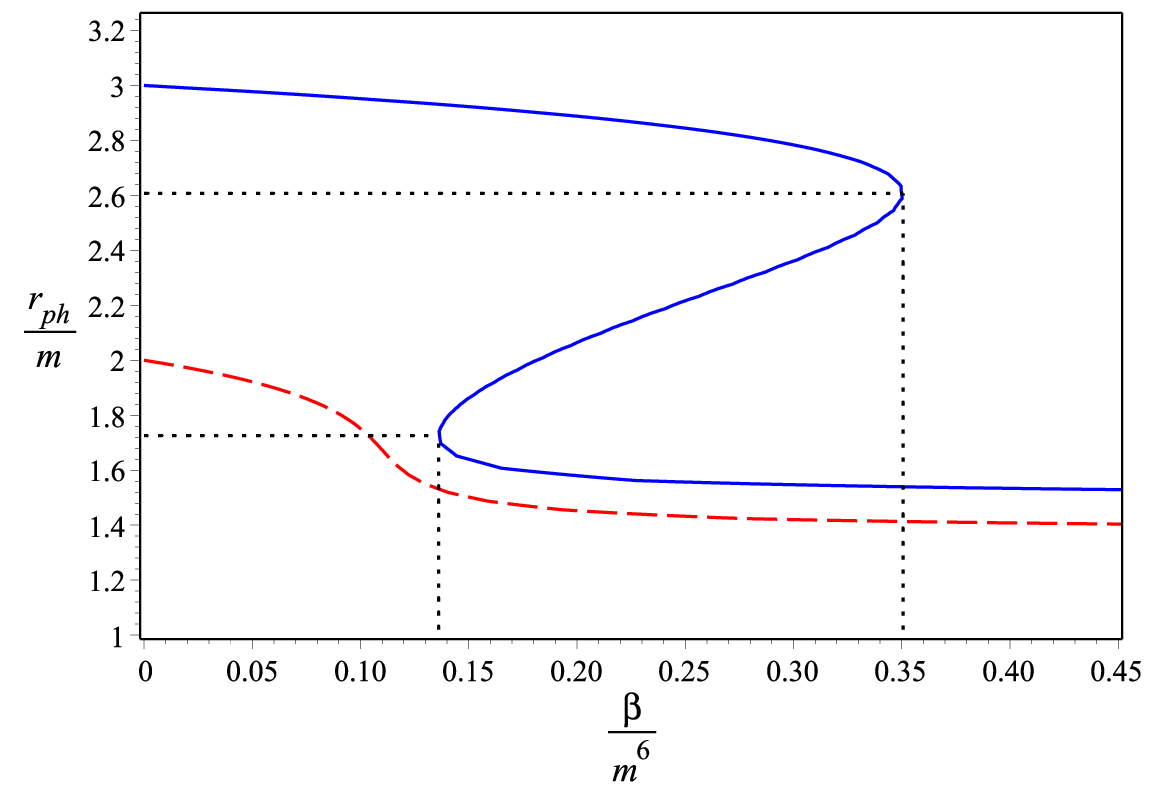}~~~~~~~~~~
	\includegraphics[width=0.75\columnwidth]{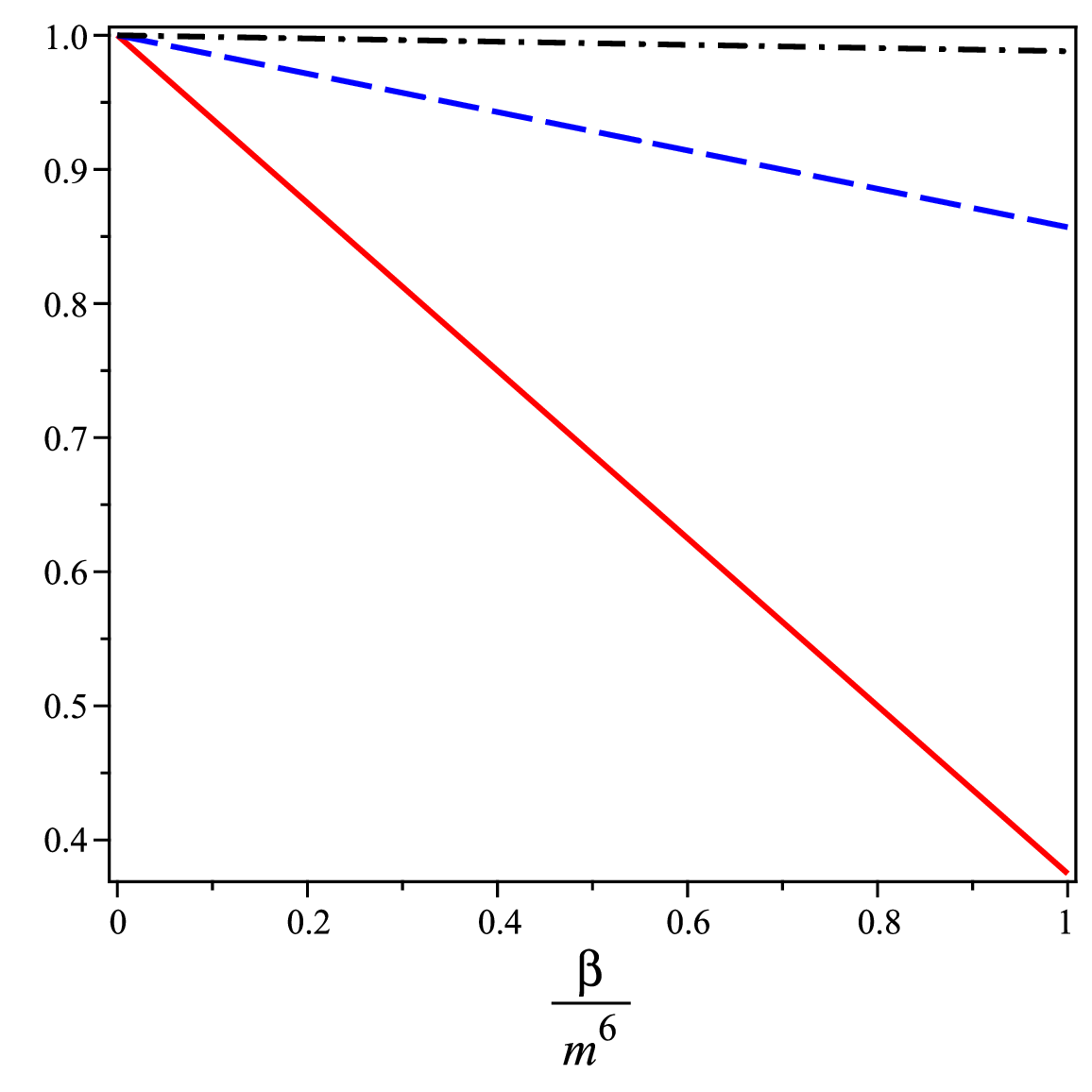}
	\caption{Left: The behavior of $\frac{r_{ph}}{m}$ (solid blue line) and $\frac{r_{+}}{m}$ (dash red line) in terms of $\beta/m^6$ for photons. Right: The behavior of $r_{+}/2m$ (solid red line), $r_{ph}/3m$ (dashed blue line) and $r_{isco}/6m$ (dash-dotted black line) in terms of $\beta/6m$ have been shown.}
	\label{riplotnull}
\end{figure*}

We now consider the behavior of null geodesics near the black hole. The geodesics equation for the case of light-like ($\mu=0$) is given by
\begin{equation}\label{geod1}
\dfrac{1}{2}\dfrac{h}{f}\dot{r}^{2}+\dfrac{1}{2}\left( \dfrac{h(r) L^{2}}{r^{2}}\right)=\dfrac{E}{2}.
\end{equation}
The effective potential is explicitly given by
\begin{equation}
    V_{eff}=\dfrac{hL^2}{r^2}=\dfrac{L^2}{r^2}-\dfrac{2L^2m}{r^3}+\beta\left(\dfrac{1024L^2m^3}{r^{11}}-\dfrac{1408L^2m^4}{r^{12}}
    \right).
\end{equation}
The EBR term includes two positive and negative terms, leading to extra extremum in the effective potential curve.
In Fig.~\ref{veffnull}, the behavior of effective potential has been shown. As expected, for $\beta\neq 0$ there are additional maximums and minimums compared to the GR in a smaller radius. 
This means there is an extra unstable circular orbit for photons for $r<3m$.
Fig.~\ref{riplotnull} displays the behavior of extremals of the effective potential in terms of $\beta/m^6$ (blue dashed line). For $0<\beta/m^6<0.136$, the radius of the photon sphere i.e., $V'_{eff}(r_{ph})=0$, is between $2.92m<r_{ph}<3m$. For $0.136<\beta/m^6<0.351$, there are three branches of extremums for which the larger $2.608m<r_{ph}<2.92m$ and smaller $1.55m<r_{ph}<1.73m$ radii that are maximums of potential correspond to unstable circular orbits for photons and the middle branch with $1.73m<r_{ph}<2.608m$ is the minimum of the potential corresponds to the stable circular orbits for photons. For $0.351<\beta/m^6$, there is an unstable circular orbit for photons in a smaller radius $1.55m<r_{ph}$. {Unlike Einstein's gravity, there is a stable circular orbit for photons.}

The radius of the unstable circular orbit (photon sphere) is given by

\begin{equation}\label{rphoton}
    r_{ph}=3m-\dfrac{2816\beta}{6561m^{5}}+\mathcal{O}(\beta^2).
\end{equation}
As shown in Fig.~\ref{riplotnull}, the photon-sphere radius becomes smaller by increasing the coupling constant of the theory. The height of the maximum is 
\begin{equation}
    V^{max}_{eff}=\dfrac{L^2}{27m^2}+\dfrac{1664L^2\beta}{531441m^8}+\mathcal{O}(\beta^2).
\end{equation}
It is clear that the effective potential increases with $\beta$.
From Fig.~\ref{veffnull}, if the energy of the non-radial ray is larger than $V_{eff}^{max}$, i.e. ($E>V_{eff}^{max}$), the incoming photon enters $r=2m$ and is captured by the black hole.
For $E<V_{eff}^{max}$, the incoming photon is scattered by the potential back to infinity. 
From Fig.~\ref{veffnull}, if the energy of the non-radial ray is larger than $V_{eff}(r=3m)$, i.e. ($E>V_{eff}(r=3m)$), the incoming photon is scattered by the potential back to infinity. For $E=V_{eff}(r_{ph})$, the incoming photon enters an unstable circular orbit of radius $r=r_{ph}$. Therefore, the impact parameter or shadow radius is given by
\begin{equation}
 R_{sh}=3\sqrt{3}m-\dfrac{832\sqrt{3}\beta}{6561 m^5}+\mathcal{O}(\beta^2).
\end{equation}
Thus to a distant observer, the apparent radius of the black hole is $R_{sh}$, and the capture cross-section is given by
\begin{equation}\label{crosssect}
  \sigma=\pi R_{sh}^{2}=27\pi m^2-\dfrac{1664\pi \beta}{729 m^4}+\mathcal{O}(\beta^2).  
\end{equation}
The EBR black hole will be seen as a black disc with the area decreasing to Einstein's gravity. Therefore, we find that the photon sphere radius of the EBR black hole exists at a lower radius but higher energy when compared with GR. 

\begin{figure*}
	\centering
\includegraphics[width=1\columnwidth]{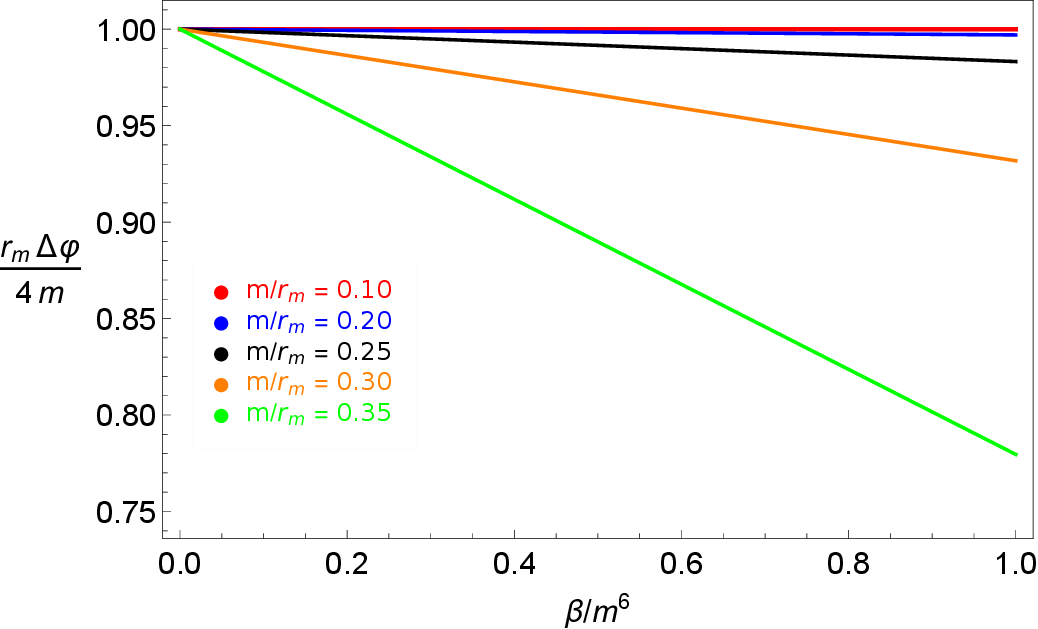}~~~~~~~~~~
\includegraphics[width=0.95\columnwidth]{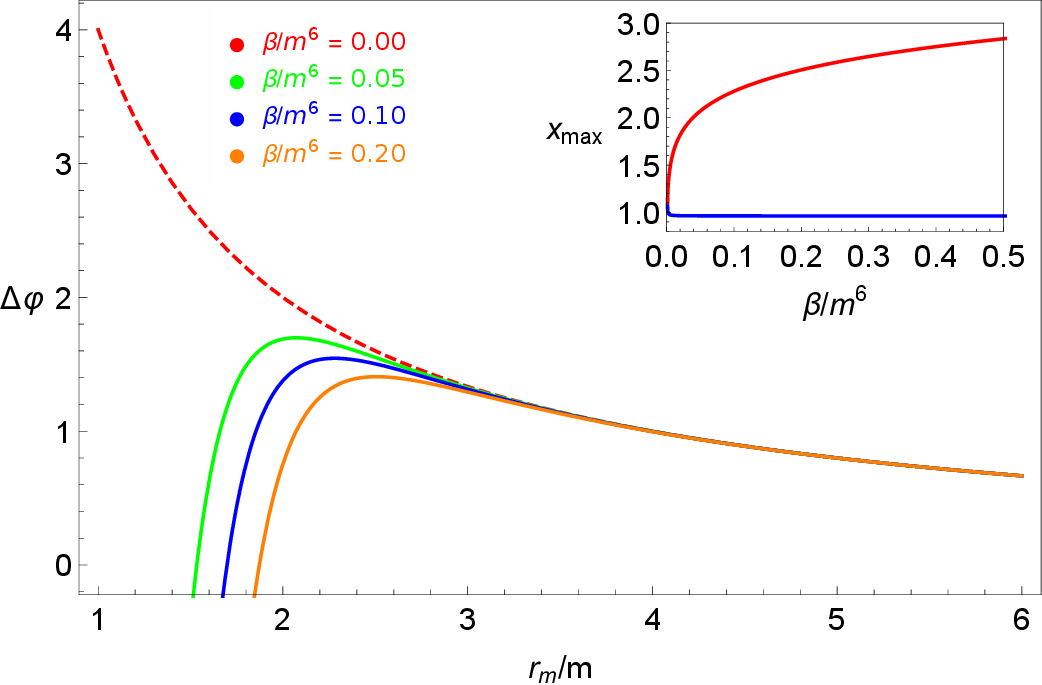}
	\caption{Left: The behavior of $r_{m}\Delta \phi/4m$ in terms of $\beta/m^6$ for $m/r_{m}=0.1,0.2,0.25,0.3,0.35$. Right: The behavior of $\Delta \phi$ in terms of $r_{m}/m$ for $\beta/m^6=0,0.05,0.1,0.2$. In the inset the behavior of $x_{max}=\left(\frac{r_{m}}{m}\right)_{max}$ in terms of $\beta/m^6$ has been shown. }
	\label{deltaphip}
\end{figure*}

The deflection of the photon as it moves from infinity to $r_m$ and off to infinity for the metric \eqref{sssmet} can be expressed as
\begin{equation}\label{eq37}
\Delta \varphi=\int_{r_{m}}^{\infty}\dfrac{2dr}{\sqrt{\dfrac{f}{h}\dfrac{r^{4}}{b^{2}}-f r^{2}}}-{\pi}=2I-\pi ,
\end{equation}
where $ b=\sqrt{\frac{r_{m}^{2}}{h(r_{m})}} $ is the impact parameter of the null ray and $ r_{m} $ is coordinate
distance of closest approach. Here $\pi$ is the change in the angle $ \varphi $ for straight line motion and is therefore subtracted out. Writing the term in the denominator
of \eqref{eq37} as  ${f(r)} r^2 \left(r^2/b^2 h(r)-1\right)$, we have
\begin{widetext}
\begin{align}
\frac{h(r_{m})}{h(r)}\dfrac{r^{2}}{r_{m}^{2}}-1 &=\left[\dfrac{1-\dfrac{2m}{r_{m}}+\left(\dfrac{1024m^3}{r_{m}^9}-\dfrac{1408m^4}{r_{m}^{10}}\right)\beta}{1-\dfrac{2m}{r}+\left(\dfrac{1024m^3}{r^9}-\dfrac{1408m^4}{r^{10}}\right)\beta}\right]\left(\dfrac{r^{2}}{r_{m}^{2}} \right)-1, \nonumber \\
&=\left(\dfrac{r}{r_{m}}\right)^{2}\left[1+2m\left(\dfrac{1}{r}-\dfrac{1}{r_{m}} \right)+1024\beta m^3 \left(\dfrac{1}{r_{m}^9}-\dfrac{1}{r^9}\right)+1408\beta m^4\left(\dfrac{1}{r^{10}}-\dfrac{1}{r_{m}^{10}}\right)\right]-1, \nonumber\\
&=\left(\dfrac{r^{2}}{r_{m}^{2}}-1\right)\left[1-\dfrac{2mr}{r_{m}(r+r_{m})}+\dfrac{1024\beta m^3}{r^{7}r_{m}^{9}}\dfrac{r^9-r_{m}^9}{r^2-r_{m}^2}+\dfrac{1408\beta m^4}{r_{m}^{10}r^{8}}\dfrac{r_{m}^{10}-r^{10}}{r^2-r_{m}^2}\right],
\label{expand}   
\end{align}
\end{widetext}
for $m \ll r$.  Therefore, the integrand becomes
\begin{widetext}
\begin{align}
I &= \int_{r_{m}}^{\infty}\dfrac{1}{\sqrt{\Big(\dfrac{1}{r_{m}^{2}}-\dfrac{1}{r^{2}}\Big)} }\Bigg[1+\dfrac{m}{r}\Big(1+\dfrac{r^{2}}{r_{m}(r+r_{m})}\Big)-\dfrac{2304m^3\beta}{r^{9}}\Big(1+\dfrac{2}{9}\dfrac{r^2}{r_{m}^9}\dfrac{r^{9}-r_{m}^9}{r^{2}-r_{m}^{2}}\Big)\nonumber\\
&~~~~+\dfrac{4288\beta m^4}{r^{10}}\Big(1-\dfrac{11}{67}\dfrac{r^2}{r_{m}^{10}}\dfrac{r_{m}^{10}-r^{10}}{r^{2}-r_{m}^{2}}-\dfrac{8}{67}\dfrac{r^2}{r_{m}^{9}}\dfrac{r^9-r_{m}^{9}}{r^{2}-r_{m}^{2}}\Big)-\dfrac{2304\beta m^4}{r^{8}r_{m}(r+r_{m})}\Bigg]\dfrac{dr}{r^{2}},
\label{eq42}
\end{align}
\end{widetext}
upon expanding in powers of $m/r$, $m/r_m$, and $\beta/m^6$. This integration can be evaluated by substituting $\sin \theta = \frac{r_m}{r}$. Thus, we obtain
\begin{widetext}
\begin{align}
I &= \int_{0}^{\frac{\pi}{2}}d\theta \Bigg[1+\dfrac{m}{r_{m}}\Big(\sin \theta+\dfrac{1}{1+\sin\theta}\Big)-\dfrac{2304\beta m^3}{r_{m}^9}\Big(\sin^{9}\theta+\dfrac{2}{9}\dfrac{1-\sin^{9}\theta}{1-\sin^{2}\theta}\Big) \nonumber \\
&~~~~~~~+\dfrac{4288\beta m^4}{r_{m}^{10}}\Big(\sin^{10}\theta -\dfrac{11}{67}\dfrac{\sin^{10}\theta-1}{1-\sin^{2}\theta}-\dfrac{8\sin\theta}{67}\dfrac{1-\sin^{9}\theta}{1-\sin^{2}\theta}\Big)-\dfrac{2304\beta m^4\sin^{9}\theta}{r_{m}^{10}(1+\sin\theta)}\Bigg], \nonumber \\
&=\dfrac{\pi}{2}+\dfrac{2m}{r_{m}}-\dfrac{98304\beta m^3}{35r_{m}^{9}}-\dfrac{(579915\pi-2502656) \beta m^4}{280r_{m}^{10}}.
\end{align}
\end{widetext}
The deflection angle is explicitly given by 
\begin{align}\label{deflect}
\Delta \varphi &= 2I- \pi \nonumber \\
&=\dfrac{4m}{r_{m}}-\dfrac{2\beta}{m^{6}}\left[\dfrac{98304 m^9}{35r_{m}^{9}}+\dfrac{(579915\pi-2502656) m^{10}}{280r_{m}^{10}}\right].
\end{align}
We remark that this formula is valid for small $\beta$ (i.e., $ \beta \rightarrow 0 $). The behavior of $\Delta \varphi$  has been shown in Fig.~\ref{deltaphip}. As can be seen, by increasing both $\beta$ and $r_{m}$ the deflection angle decreases. In the right panel, we can see the deflection angle has a maximum angle while in Einstein's gravity deflection angle decreases monotonically (red dashed curve). As the subplot has shown the maximum deflection angle increases as the coupling $\beta$ increases (red curve in the inset). 

We next consider the shadow of these black holes. We follow up the null geodesics that satisfy the condition $ V_{eff}^{''}<0 $, i.e., unstable circular orbits.  The angular radius ($\Gamma$) of the shadow as seen by an
observer at $ r_{0} $ is \cite{Hennigar:2018hza} 
\begin{equation}
\sin^{2}(\Gamma)=\dfrac{r_{ph}^{2} h(r_{0})}{r_{0}^{2} h(r_{ph})}.
\end{equation}
Using (\ref{eqapproximat}) for $ h(r) $, upon expanding in powers of $m/r_{0}$, $m/r_{ph}$, and $\beta/m^6$ we obtain
\begin{widetext}
\begin{align}
\dfrac{r_{ph}^{2}h(r_{0})}{r_{0}^{2}h(r_{ph})} &=\dfrac{r_{ph}^{2}}{r_{0}^{2}}\left[\dfrac{1-\dfrac{2m}{r_{0}}+\left(\dfrac{1024m^3}{r_{0}^{9}}-\dfrac{1408m^4}{r_{0}^{10}}\right)\beta}{1-\dfrac{2m}{r_{ph}}+\left(\dfrac{1024m^3}{r_{ph}^{9}}-\dfrac{1408m^4}{r_{ph}^{10}}\right)\beta} \right],
\nonumber\\
&= \dfrac{r_{ph}^{2}}{r_{0}^{2}}\left[\left( 1-\dfrac{2m}{r_{0}}+\left(\dfrac{1024m^3}{r_{0}^{9}}-\dfrac{1408m^4}{r_{0}^{10}}\right)\beta\right)\left(1+\dfrac{2m}{r_{ph}}-\left(\dfrac{1024m^3}{r_{ph}^{9}}-\dfrac{1408m^4}{r_{ph}^{10}}\right)\beta\right)\right],\nonumber\\
&= \dfrac{r_{ph}^{2}}{r_{0}^{2}}\left[1+2m\left(\dfrac{1}{r_{ph}}-\dfrac{1}{r_{0}}\right)+1024\beta m^{3}\left(\dfrac{1}{r_{0}^{9}}-\dfrac{1}{r_{ph}^9}\right)+1408\beta m^4\left(\dfrac{1}{r_{ph}^{10}}-\dfrac{1}{r_{0}^{10}}\right) \right]. 
\end{align}
\end{widetext}
Thus, it yields
\begin{align}
\sin(\Gamma)&=\dfrac{r_{ph}}{r_{0}}+\dfrac{m(r_{0}-r_{ph})}{r_{0}^{2}}+\nonumber\\
&\dfrac{\beta}{m^{6}}\left[512 m^9\left(\dfrac{r_{ph}^9-r_{0}^{9}}{r_{0}^{10}r_{ph}^{8}}\right)+704 m^{10}\left(\dfrac{r_{0}^{10}-r_{ph}^{10}}{r_{0}^{11}r_{ph}^{9}}\right)\right], \label{sineGamma}
\end{align}
where $ r_{ph} $ is the radius of the photon sphere and $ \Gamma $ is the angle subtended by the shadow's radius as seen by an observer at $r_{0}$.

\begin{figure*}
	\centering
\includegraphics[width=0.95\columnwidth]{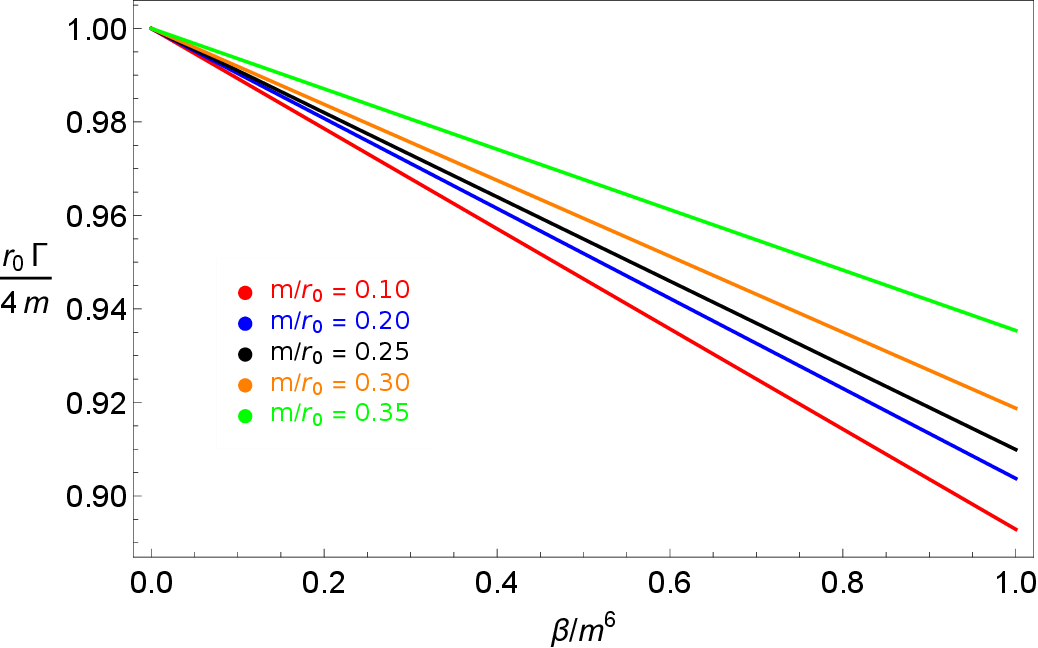}~~~~~~~~~~	\includegraphics[width=0.95\columnwidth]{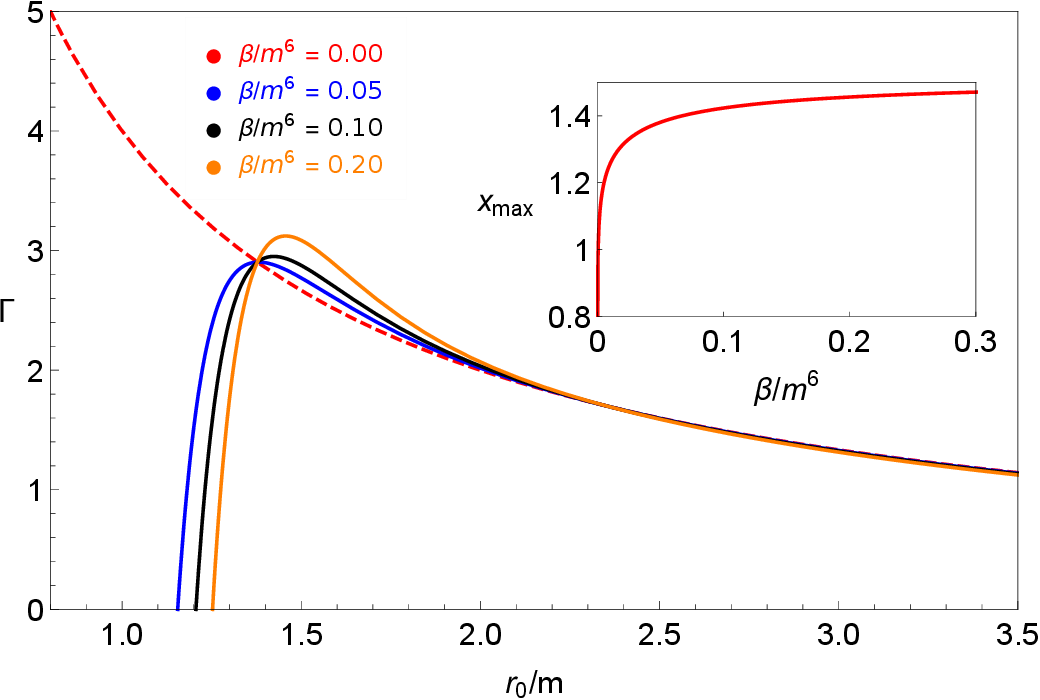}
\caption{Left: The behavior of $r_{0}\Gamma/4m$ in terms of $\beta/m^6$ for $m/r_{0}=0.1,0.2,0.25,0.3,0.35$. Right: The behavior of $\Gamma$ against $r_{0}/m$ for $\beta/m^6=0,0.05,0.1,0.2$. In the inset the behavior of $x_{max}=\left(\frac{r_{0}}{m}\right)_{max}$ in terms of $\beta/m^6$ has been shown.}
	\label{deltaphipp}
\end{figure*}

In the case of small $ \Gamma $, we have $\sin(\Gamma)\approx \Gamma$, so
\begin{equation}
\Gamma=\Gamma_{Ein}+\Gamma_{BR}
\end{equation}
Therefore, the first two terms on the right-hand side of \eqref{sineGamma} are the Einstein term ($\Gamma_{Ein}$) while the $\beta-$dependence term can be identified as the correction term coming from the BR corrections. Then, we substitute the unstable circular orbit of photon sphere \eqref{rphoton} to obtain 
\begin{align}
    \Gamma &= \dfrac {4m}{r_{0}}+\dfrac{\beta}{m^{6}}\Big(-\dfrac{9280m}{19683r_{0}}+\dfrac{2816m^2}{6561r_{0}^{2}}+\dfrac{1536m^{10}}{r_{0}^{10}}\nonumber\\
    &~~~~-\dfrac{2112m^{11}}{r_{0}^{11}}\Big)+\mathcal{O}(\beta^2)
\end{align}
In Fig.~\ref{deltaphipp}, the behavior of $\Gamma$ has been shown. As can be seen, the radius of the shadow of the EBR theory is greater than Einstein's gravity, and by increasing the coupling of theory the radius of the shadow decreases. In the right panel, the radius of the shadow has a maximum of Einstein gravity, and increasing the coupling constant increases the curve's maximum.

Finally, we consider Shapiro's time delay to obtain a bound on the coupling constant $\beta$. The general expression for time delay for the metric (\ref{sssmet}) is  
\begin{align}
t(r_{0},r)&=\int_{r_{0}}^{r}\dfrac{dr}{\sqrt{f(r)h(r)\left( 1-\dfrac{r_{0}^{2}}{r^{2}}\dfrac{h(r)}{h(r_{0})}\right) }},\nonumber \\
&\equiv \int_{r_{0}}^{r} \mathcal{T} dr.
\end{align}
To evaluate the integral, we expand the metric at the asymptotic regime. Similar manipulations as before yield the integrand to be
\begin{widetext}
\begin{align}
\mathcal{T} &=\dfrac{1}{\sqrt{1-\dfrac{r_{0}^{2}}{r^{2}}}}\Bigg[1+\dfrac{m}{r}\Big( 1+\dfrac{r_{0}}{(r+r_{0})}\Big)-\dfrac{2816\beta m^3}{r^9}\Big(1-\dfrac{2}{11}\dfrac{r_{0}^{9}-r^9}{r_{0}^7(r^2-r_{0}^{2})}\Big) +\dfrac{10624\beta m^4}{r^{10}}\Big(1-\dfrac{22}{83}\dfrac{r_{0}}{r_{0}+r} \nonumber\\
&~~~~+\dfrac{8}{83}\dfrac{r_{0}^{9}-r^9}{r_{0}^7(r^2-r_{0}^2)}+\dfrac{11}{166}\dfrac{r^{10}-r_{0}^{10}}{r_{0}^{8}(r^2-r_{0}^2)}\Big)\Bigg].
\end{align}
\end{widetext}
Now, integral is elementary and we find that the time required for light to move from $ r_{0} $ to $ r $ is
\begin{widetext}
\begin{align}
t(r,r_{0})&\approx \sqrt{r^{2}-r_{0}^{2}}+m\ln\Big(\dfrac{r+\sqrt{r^{2}-r_{0}^{2}}}{r_{0}}\Big)+m\sqrt{\dfrac{r-r_{0}}{r+r_{0}}}-\dfrac{2816\beta m^3}{35r_{0}^{8}}\sqrt{1-\dfrac{r_{0}^{2}}{r^{2}}}\Big(16+\dfrac{8r_{0}^{2}}{r^2}\nonumber\\
&+\dfrac{6r_{0}^{4}}{r^{4}}+\dfrac{5r_{0}^{6}}{r^{6}}\Big)-\dfrac{512\beta m^{3}}{35r_{0}^{8}}\sqrt{\dfrac{r-r_{0}}{r+r_{0}}}\Big(128+\dfrac{93r_{0}}{r}+\dfrac{29r_{0}^2}{r^2}+\dfrac{29r_{0}^{3}}{r^{3}}+\dfrac{13r_{0}^{4}}{r^{4}}+\dfrac{13r_{0}^{5}}{r^{5}}+\dfrac{5r_{0}^{6}}{r^{6}}\nonumber\\
&+\dfrac{5r_{0}^{7}}{r^{7}}\Big)+\dfrac{83\beta m^4}{6r_{0}^{9}}\Big(105\pi+\dfrac{210r_{0}}{r}\sqrt{1-\dfrac{r_{0}^{2}}{r^{2}}}+\dfrac{140r_{0}^{3}}{r^{3}}\sqrt{1-\dfrac{r_{0}^{2}}{r^{2}}}+\dfrac{112r_{0}^{5}}{r^{5}}\sqrt{1-\dfrac{r_{0}^{2}}{r^{2}}}+\dfrac{96r_{0}^{7}}{r^{7}}\nonumber\\
&\sqrt{1-\dfrac{r_{0}^{2}}{r^{2}}}-210\arctan\Big({\Big(\frac{r^2}{r_{0}^2}-1\Big)}^{-\frac{1}{2}}\Big)\Big)+...
\end{align}
\end{widetext}
Working in the asymptotic regime schematically this expression is
 \begin{align}
t(r,r_{0})&=t_{SR}(r,r_{0})+\Delta t_{GR}(r,r_{0})+\Delta t_{BR},(r,r_{0})\nonumber\\
&=t_{SR}(r,r_{0})+\Delta t(r,r_{0}),
\end{align}
where $ t_{SR}=\sqrt{r^{2}-r_{0}^{2}} $ is the special relativistic contribution of the propagation of light in
flat spacetime.
So, the maximum round-trip excess time delay is given by
\begin{widetext}
\begin{align}\label{eqshap}
\Delta t(r,r_{0})&=4m\ln\Big(\dfrac{r+\sqrt{r^2-r_0^2}}{r_0}\Big)+4m\sqrt{\dfrac{r-r_{0}}{r+r_{0}}}-\dfrac{11264\beta m^3}{35r_{0}^{8}}\sqrt{1-\dfrac{r_{0}^{2}}{r^{2}}}\Big(16+\dfrac{8r_{0}^{2}}{r^2}+\dfrac{6r_{0}^{4}}{r^{4}}+\nonumber\\
&\dfrac{5r_{0}^{6}}{r^{6}}\Big)-\dfrac{2048\beta m^{3}}{35r_{0}^{8}}\sqrt{\dfrac{r-r_{0}}{r+r_{0}}}\Big(128+\dfrac{93r_{0}}{r}+\dfrac{29r_{0}^2}{r^2}+\dfrac{29r_{0}^{3}}{r^{3}}+\dfrac{13r_{0}^{4}}{r^{4}}+\dfrac{13r_{0}^{5}}{r^{5}}+\dfrac{5r_{0}^{6}}{r^{6}}+\dfrac{5r_{0}^{7}}{r^{7}}\Big)\nonumber\\
&+\dfrac{332\beta m^4}{6r_{0}^{9}}\Big(105\pi+\dfrac{210r_{0}}{r}\sqrt{1-\dfrac{r_{0}^{2}}{r^{2}}}+\dfrac{140r_{0}^{3}}{r^{3}}\sqrt{1-\dfrac{r_{0}^{2}}{r^{2}}}+\dfrac{112r_{0}^{5}}{r^{5}}\sqrt{1-\dfrac{r_{0}^{2}}{r^{2}}}+\dfrac{96r_{0}^{7}}{r^{7}}\sqrt{1-\dfrac{r_{0}^{2}}{r^{2}}}\nonumber\\
&-210 \arctan\Big({\Big(\frac{r^2}{r_{0}^2}-1\Big)}^{-\frac{1}{2}}\Big)\Big)+...
\end{align}
\end{widetext}
We have partitioned the expression into the general relativistic (GR) and the BR corrections. Here  $ r_{0} $ is the distance of the closest approach of the radar wave to the center of the Sun, $ r $ is the distance along the line of light traveling from the Earth to the point of closest approach to the Sun and $ r\gg r_{0}$.

Deviations of this result from the prediction of general relativity have been constrained to be less than $0.000012$.  A careful numerical evaluation of the integrals reveals that it provides a numerical constraint on the coupling constant $\beta$ as follows
\begin{equation}
  \dfrac{\Delta t_{BR}}{\Delta t_{GR}}<0.000012\;\;\;\to\;\;\; \beta<0.4\times 10^{39} m_{\odot}^{6},
\end{equation}
Here we have used $r=10^{11}m$, $r_{0}=10^{8}m$ and $m_{\odot}=1477m$.

\section{Dynamical Stability and 
Evolution}\label{dynsta}

Here, we analyze the behavior of massless scalar and electromagnetic perturbations within the spacetime of the black hole under study, adopting the assumption that the test field has a minimal impact on the black hole’s overall spacetime structure. This simplification enables us to examine the intrinsic properties of the perturbations while maintaining mathematical consistency with the black hole spacetime framework. The analysis begins by deriving Schr\"odinger-like wave equations, which, in the case of massless scalar fields, take on a Klein-Gordon form. This transformation aligns the perturbative framework with the spacetime under study and allows for a focused exploration of QNMs, the characteristic oscillations that arise when a black hole is perturbed.
To compute these QNMs, we employ the Wentzel–Kramers–Brillouin (WKB) approximation up to the sixth order, a semi-classical method suitable for examining wave equations in a curved spacetime. This approach allows for a detailed analysis of the potential barrier created by perturbations near the black hole and the behavior of the scalar field within this region. From this, we calculate the frequencies and damping rates of the QNMs, providing insights into the stability and dynamic response of the black hole to external disturbances. WKB has been widely used in various studies on QNMs of black holes  \cite{Schutz:1985km,PhysRevD.35.3621,konoplya2003quasinormal,zhidenko2003quasi,Burikham:2017gdm,Ponglertsakul:2018smo,matyjasek2019quasinormal,konoplya2019higher,Santos:2019yzk,Ponglertsakul:2020ufm,Ponglertsakul:2022vni,gogoi2023quasinormal,tangphati2024magnetically,gogoi2024constraints, Gogoi:2024epx, Gogoi:2024scc}.

The metric under axial perturbations is represented as \cite{Bouhmadi-Lopez:2020oia}:
\begin{align}
    \label{pert_metric}
ds^2 &=  r^2 \sin^2\!\theta\, (d\phi - p_2(t,r,\theta)\, dr -  p_1(t,r,\theta)\,
dt \notag \\ &- p_3(t,r,\theta)\, d\theta)^2 + g_{rr}\, dr^2 -\, |g_{tt}|\, dt^2 +
r^2 d\theta^2,
\end{align}
where the parameters $p_1$, $p_2$, and $p_3$ reflect perturbations to the black hole spacetime, while the standard metric functions $g_{tt}$ and $g_{rr}$ represent the unperturbed terms. This framework provides a foundation for studying the stability and response of black holes under external influences, contributing to our understanding of black hole dynamics.

To explore the behavior of the massless scalar field in the vicinity of the black hole, we assume that the scalar field has a negligible impact on the surrounding spacetime, allowing us to focus purely on the perturbative aspects. This assumption enables us to simplify the perturbed metric $ ds^2 $ in \eqref{pert_metric} to 
\begin{equation}
ds^2 = g_{rr}\, dr^2 +r^2 d \Omega^2 - |g_{tt}|\, dt^2,
\end{equation}
where $ g_{rr} $ and $ g_{tt} $ represent the radial and temporal components of the metric, respectively. This reduced form of the metric allows for a straightforward application of the Klein-Gordon equation in curved spacetime, as described by:
\begin{equation}
\square \Phi = \frac{1}{\sqrt{-g}} \partial_\mu (\sqrt{-g} g^{\mu\nu} \partial_\nu \Phi) = 0.
\end{equation}
The Klein-Gordon equation, which governs the behavior of scalar fields in curved spacetime, is particularly useful for investigating quasinormal modes associated with scalar perturbations.

We proceed by decomposing the scalar field $ \Phi(t,r,\theta,\phi) $ in terms of spherical harmonics, as shown:
\begin{equation}
\Phi(t,r,\theta, \phi) = \frac{1}{r} \sum_{l,m} \psi_l(t,r) Y_{lm}(\theta, \phi),
\end{equation}
where $ Y_{lm}(\theta, \phi) $ represents the spherical harmonics, with $ l $ and $ m $ as angular momentum quantum numbers. The function $ \psi_l(t,r) $ encapsulates the radial and temporal dependence of the wave function. By substituting this expression into the Klein-Gordon equation, we derive a second-order differential equation for the radial component of the scalar field, yielding the Schrödinger-like form:
\begin{equation}
\partial^2_{r_*} \psi(r_*)_l + \omega^2 \psi(r_*)_l = V_s(r) \psi(r_*)_l,
\end{equation}
where $ r_* $ denotes the tortoise coordinate, defined in such a way as to map the spatial variable $ r $ into a coordinate that smooths out the radial infinity at the black hole horizon:
\begin{equation}
\frac{dr_*}{dr} = \sqrt{g_{rr}\, |g_{tt}^{-1}|}.
\end{equation}
The transformation to the tortoise coordinate is crucial for handling the asymptotic behavior of the field at the horizon, making it well-suited for analyzing quasinormal modes.

The effective potential $ V_s(r) $ for the scalar field in this black hole spacetime is given by:
\begin{equation}
V_s(r) = |g_{tt}| \left( \frac{l(l+1)}{r^2} + \frac{1}{r \sqrt{|g_{tt}| g_{rr}}} \frac{d}{dr} \sqrt{|g_{tt}| g_{rr}^{-1}} \right),
\end{equation}
where the angular quantum number $ l $ represents the multipole moment of the quasinormal modes. This potential acts as a barrier that affects the propagation of the scalar field, with its height and shape depending on the metric components and the angular momentum of the mode. The effective potential is central to understanding the stability of the quasinormal modes, as its peaks dictate where the field experiences the strongest oscillations. Higher multipole values result in increased potential barriers, confining the perturbations closer to the black hole.

Similarly, in the case of electromagnetic perturbation, the associated potential can be given by the following expression:
\begin{equation}\label{Ve}
V_e(r) = |g_{tt}|\, \dfrac{l(l+1)}{r^2}. 
\end{equation}

Although our prime focus is on scalar perturbation, we shall compare the QNMs from both types of perturbations in order to get a comparative idea regarding the variations w.r.t. the parameter space of the black hole system.

\begin{figure}
	\centering
\includegraphics[width=0.9\columnwidth]{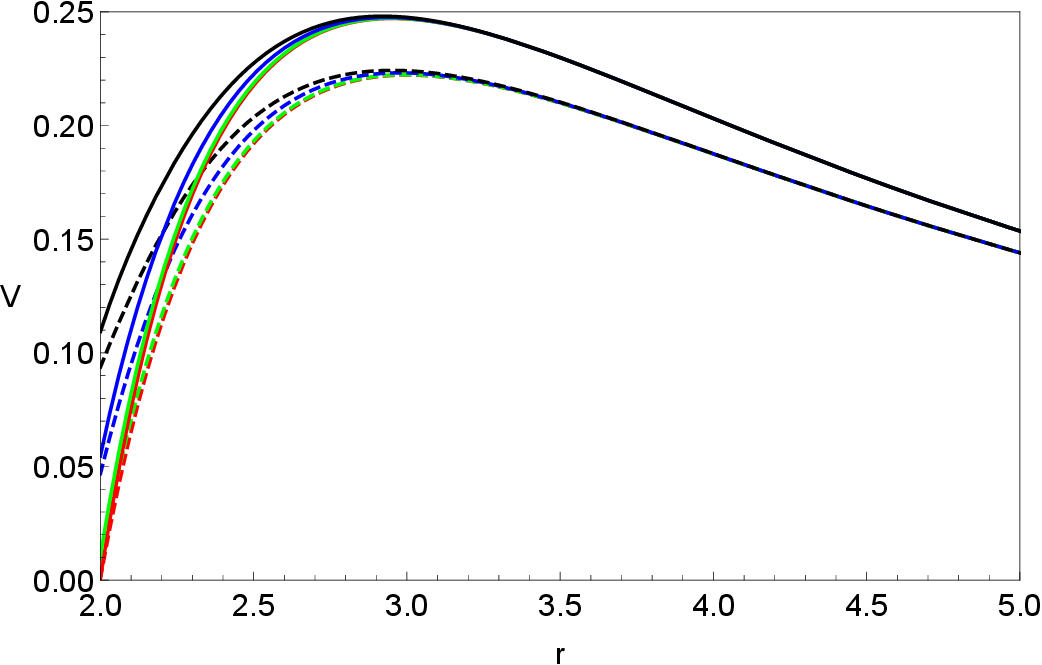}\\
\includegraphics[width=0.9\columnwidth]{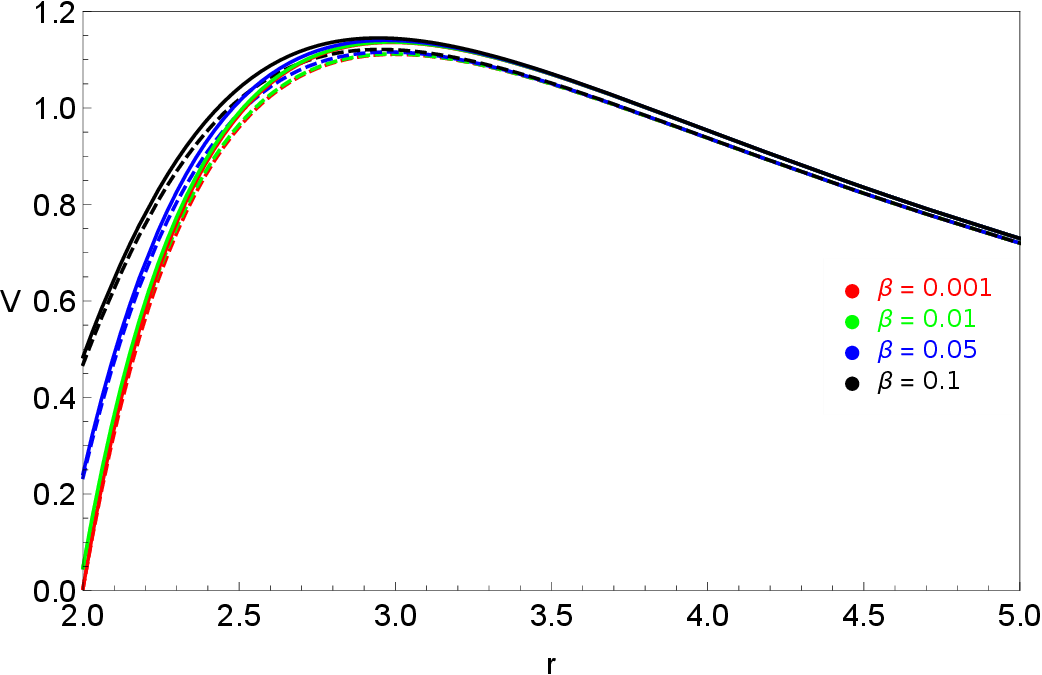}
	\caption{Effective potential $V_s$ (solid lines) and $V_e$ (dashed lines) as a function of $r$ for $m=1$ and varying $\beta$. Top: $l=2$ Bottom: $l=5$.}
	\label{fig:veffqnm}
\end{figure}

As a demonstration, the effective potential of scalar and electromagnetic perturbation up to linear in $\beta$ are plotted in Fig.~\ref{fig:veffqnm}. It is clear that the magnitude of $V_s$ is greater than $V_e$. The difference becomes less obvious as $l$ increases. Moreover, increasing $\beta$ leads to an increasing in the magnitude of the effective potential. This can be seen easily for small $r$. However, as $r$ increases, this trend is less evident. This is expected since the $\beta-$dependence term is heavily suppressed at large $r$.

By examining the quasinormal mode frequencies and damping rates derived from this framework, we gain insights into the stability and resonance characteristics of the black hole under scalar perturbations. This analysis provides a deeper understanding of the black hole’s response to external disturbances, shedding light on its inherent stability. Furthermore, the effective potential profile $ V_s(r) $ can serve as an indicator for identifying distinct characteristics of black holes in various modified theories, providing a foundation for observational tests through phenomena such as gravitational waves generated by perturbative events near the black hole.
 The QNMs in the large $l$ is given by 
\begin{equation}
    \omega^2=V_{s}(r_{p})-i\left(n+\dfrac{1}{2}\right)\sqrt{-2\dfrac{d^{2}V_{s}(r_{p})}{dr_*^2}},
\end{equation}
here $r_{p}$ is the photon sphere radius obtained in equation \eqref{rphoton}.
Therefore, the quasi-normal frequencies are given as
\begin{align} \label{qnm_large_l}
\omega=&\omega_{r}+i\omega_{i}=\left(l+\dfrac{1}{2}\right)\left[\dfrac{\sqrt{3}}{9m}+\dfrac{832\sqrt{3}}{177147m^7}\beta\right]\nonumber\\
    &-i\left(n+\dfrac{1}{2}\right)\left[\dfrac{\sqrt{3}}{9m}-\dfrac{3200\sqrt{3}}{177147m^7}\beta\right]+\mathcal{O}(\beta^2).
\end{align}
It is easy to show that the real part of QNM is inversely proportional to the shadow radius $w_{r}\approx 1/R_{sh} $.\\

For smaller values of the multipole moment $l$, the above approximation does not hold well. Hence to have an idea about the variation of QNMs for comparatively smaller $l$, we utilize the 6th-order WKB approximation method with Pad\'e averaging. This method is an effective tool for calculating QNMs numerically when the overtone number $n$ is smaller than the multipole moment $l$ \cite{Schutz:1985km, PhysRevD.35.3621, konoplya2003quasinormal, Matyjasek:2017psv, matyjasek2019quasinormal,konoplya2019higher}. One should note that the accuracy of the method increases with an increase in the value of $l-n$, where $n$ is the overtone number.

In the higher-order WKB approximation, the oscillation frequency $\omega$ of ring-down gravitational waves is given by:  
\begin{equation}
\omega = \sqrt{-\, i \left[ (n + 1/2) + \sum_{k=2}^{\bar{n}} \bar{\Lambda}_k \right] \sqrt{-2 V_0''} + V_0},
\end{equation} 
where $\bar{n}$ denotes the order of the WKB method.  
The parameter $n$, (overtone number) assumes discrete values starting from zero $( i.e., n = 0, 1, 2, \ldots)$. The term $V_0$ represents the value of the potential function $V(r)$ at the radial coordinate $r_{\text{max}}$, which corresponds to the maximum of the potential. Similarly, $V_0''$ is the second derivative of $V(r)$ with respect to $r$, evaluated at $r_{\text{max}}$. The coordinate $r_{\text{max}}$, being the location of the potential's peak, plays a crucial role in the analysis.

\begin{table}[ht]
\caption{The scalar QNMs for $n= 0$, $m=1$, $k = 1$ and $\beta=0.3$ using the 6th order WKB approximation method averaged with  Pad\'e approximants.}
\label{tab01}
\begin{center}
{\small 
\begin{tabular}{|cccc|}
\hline
\;\;$l$ &  \;\; Pad\'e averaged WKB\;\;
& $\Delta_{rms}$ & $\Delta_6$ \\ \hline
 1 & $0.260079\, -0.0737376 i$ & 0.0292523 & 0.0361886 \\
 2 & $0.467267\, -0.0741977 i$ & 0.00960566 & 0.0138136 \\
 3 & $0.669718\, -0.075608 i$ & 0.00433394 & 0.00411132 \\
 4 & $0.868963\, -0.0760915 i$ & 0.00240929 & 0.0056777 \\
 5 & $1.06672\, -0.0762185 i$ & 0.00151631 & 0.00390293 \\
 6 & $1.26373\, -0.0762001 i$ & 0.00103198 & 0.00293025 \\
 7 & $1.46032\, -0.0761117 i$ & 0.000740407 & 0.00229974 \\
 8 & $1.65667\, -0.0759843 i$ & 0.00055158 & 0.00185319 \\
\hline
\end{tabular}
}
\end{center}
\end{table}

\begin{table}[ht]
\caption{The electromagnetic QNMs for $n= 0$, $m=1$, $k = 1$ and $\beta=0.3$ using the 6th order WKB approximation method averaged with  Pad\'e approximants. }
\label{tab02}
\begin{center}
{\small 
\begin{tabular}{|cccc|}
\hline
\;\;$l$ &  \;\; Pad\'e averaged WKB\;\;
& $\Delta_{rms}$ & $\Delta_6$ \\ \hline
 1 & $0.236215\, -0.0656175 i$ & 0.0200233 & 0.0241729 \\
 2 & $0.451939\, -0.0714657 i$ & 0.00701473 & 0.00821482 \\
 3 & $0.657339\, -0.073814 i$ & 0.00352421 & 0.0079358 \\
 4 & $0.85862\, -0.0748272 i$ & 0.00209403 & 0.00503046 \\
 5 & $1.0579\, -0.0752969 i$ & 0.00137158 & 0.00362057 \\
 6 & $1.25608\, -0.075506 i$ & 0.000957178 & 0.00277424 \\
 7 & $1.45359\, -0.0755735 i$ & 0.000698226 & 0.00220088 \\
 8 & $1.65067\, -0.0755566 i$ & 0.000526166 & 0.00178639 \\
\hline
\end{tabular}
}
\end{center}
\end{table}

\begin{figure*}[htbp]
\centerline{
   \includegraphics[scale = 0.5]{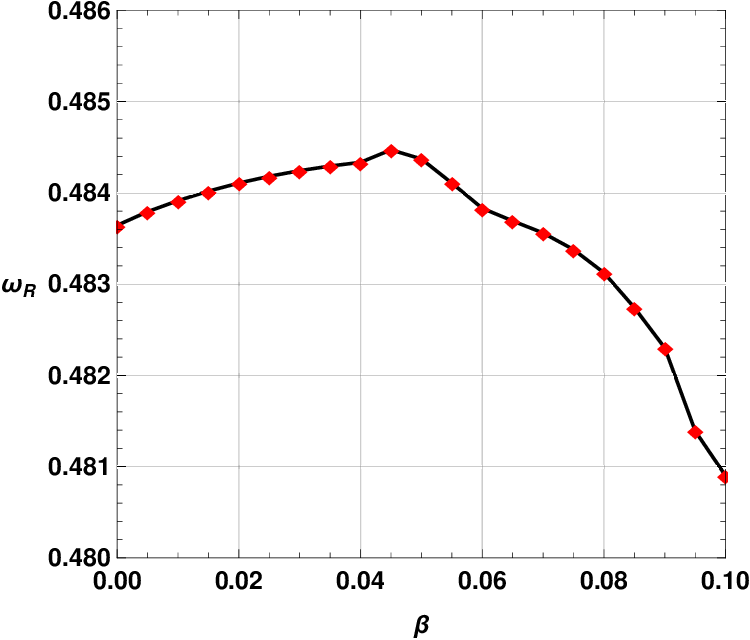}\hspace{0.5cm}
   \includegraphics[scale = 0.5]{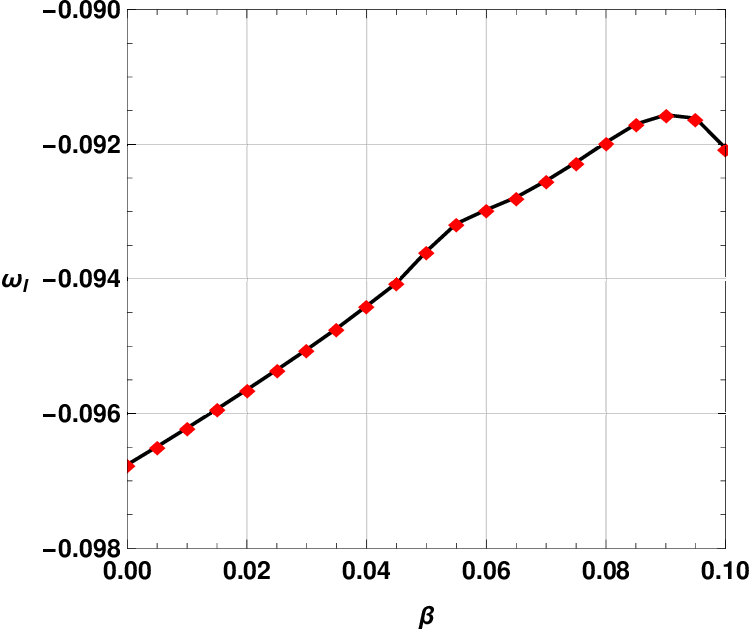}} \vspace{-0.2cm}
\caption{The real (left panel) and imaginary (right panel) parts of the QNMs for massless scalar perturbations as a function of the parameter $\beta$ with $m=1$, $n= 0$, $l=2$ and $k = 1$. In this case, we have considered 6th order Pad\'e averaged WKB approximation method and considered up to $\beta$ terms in the expansion of the potential and metric function in tortoise coordinate.}
\label{QNMs01}
\end{figure*}

\begin{figure*}[htbp]
\centerline{
   \includegraphics[scale = 0.5]{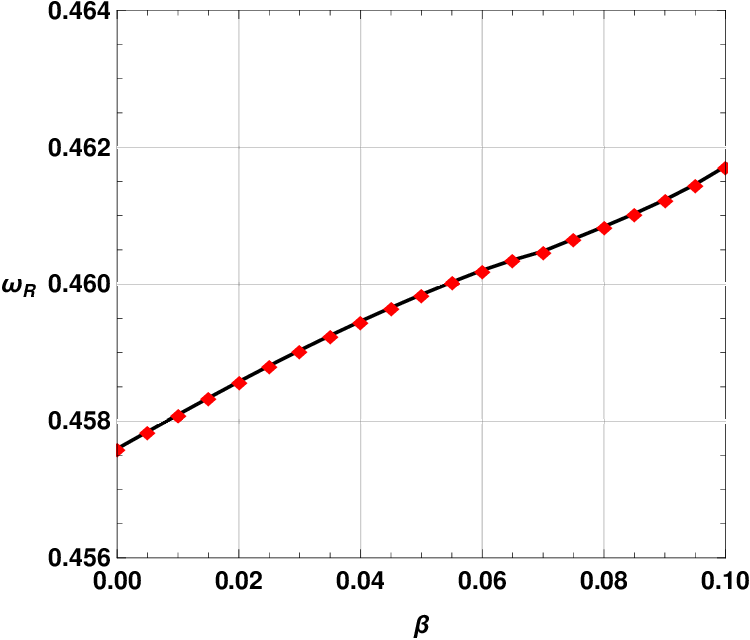}\hspace{0.5cm}
   \includegraphics[scale = 0.5]{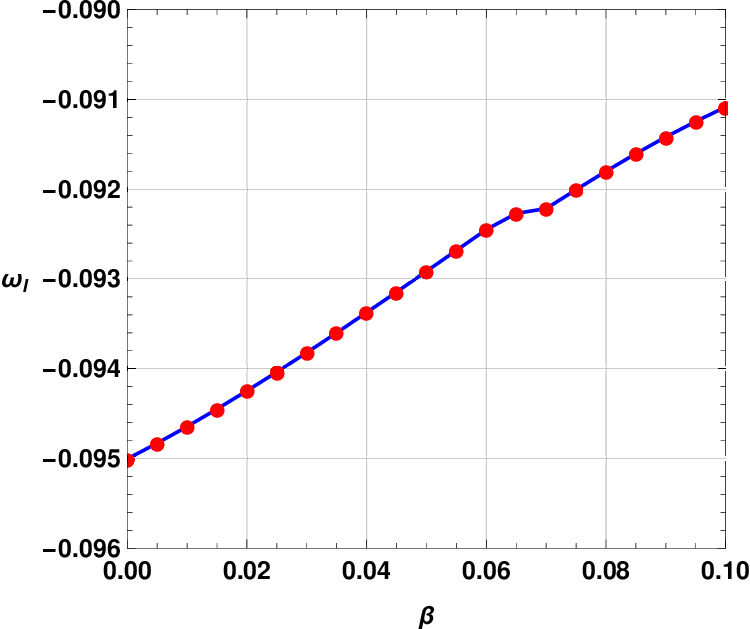}} \vspace{-0.2cm}
\caption{The real (left panel) and imaginary (right panel) parts of the QNMs for vector perturbations as a function of the parameter $\beta$ with $m=1$, $n= 0$, $l=2$ and $k = 1$. In this case, we have considered 6th order Pad\'e averaged WKB approximation method and considered up to $\beta$ terms in the expansion of the potential and metric function in tortoise coordinate.}
\label{QNMs02}
\end{figure*}

Here, we have used the Pad\'e averaged 6th-order WKB approximation method to obtain the QNMs for different model parameter values. The numerical values of QNMs for fundamental overtone are shown in Tables \ref{tab01} and \ref{tab02}. In the Tables, $\Delta_{rms}$ represents rms error associated with Pad\'e averaging. The other term in the 4th column represents error associated with the WKB method which is defined by
\cite{konoplya2019higher} 
\begin{equation}
\Delta_6 = \dfrac{\vline \; \omega_7 - \omega_5 \; \vline}{2},
\end{equation}
where the terms $\omega_7$ and $\omega_5$  denote the QNMs found using the $7$th and $5$th order Pad\'e averaged WKB method.

One may note that for this black hole metric, the errors associated with the QNMs are comparatively larger than those obtained for a standard Schwarzschild black hole. This is due to the complex nature of the black hole spacetime. However, with an increase in the value of multipole moment $l$, we observe a decline in the error values. Another observation from the tables is that the errors in the case of electromagnetic or vector perturbations are smaller.

To see the variation of QNMs spectra w.r.t. the model parameter $\beta$, we plot the real and imaginary parts of QNMs in Fig.~\ref{QNMs01} and \ref{QNMs02}. One may note that in these figures, we have neglected $\beta^2$ and higher order terms while calculating the QNMs. On the first panel of Fig.~\ref{QNMs01}, we have shown the variation of the oscillation frequency of ring-down GWs with respect to the model parameter $\beta$. It can be seen that initially, with an increase in the value of $\beta$, the oscillation frequency of GWs increases slowly. But after reaching a threshold point, oscillation frequency starts to decrease rapidly. On the other hand, in Fig.~\ref{QNMs02}, the first panel shows a different picture for the electromagnetic perturbation. In this case, with an increase in the model parameter $\beta$, the oscillation frequency increases. In the cases of both scalar and electromagnetic perturbations, the damping rate decreases with an increase in the model parameter $\beta$. However, in scalar perturbation, towards higher values of $\beta$, we observe a slight increase in the damping rate. Here we have found that for smaller values of $\beta$, the model parameter has similar impacts on the QNMs spectrum for both scalar and electromagnetic perturbations. Therefore, using classical perturbations, we discover that the black hole solutions of the theory are stable against linear scalar and electromagnetic perturbations (up to first order in $\beta$).

Here, we shall consider the evolution of the BH solution by determining its evaporation. In Hawking's picture, the time evolution of evaporation in the high-frequency limit is given as
\begin{equation}
{\frac{dm}{dt}}=-\dfrac{\pi^2}{60}(\sigma_{g}+\sigma_{\gamma}+\dfrac{21}{8}\sigma_{\nu}) T^{4}\,,
\end{equation}
where $\sigma_{i}$, are the thermally averaged cross sections of the black hole for gravitons, $g$, photons, $\gamma$, and neutrinos, $\nu$. In the following, we define
\begin{equation}
\rho=\left(\sigma_{g}+\sigma_{\gamma}+\dfrac{21}{8}\sigma_{\nu}\right)\dfrac{1}{\sigma_{0}}\,,
\end{equation}
where $\sigma_{0}$ is the geometrical optics cross section {of black hole}. The cross sections for neutrinos, photons, and gravitons can be estimated as 
\begin{equation}
{\sigma_{\nu}\sim 0.67\sigma_{0},\;\;\;\;\sigma_{\gamma}\sim 0.24\sigma_{0},\;\;\;\;\sigma_{g}\sim 0.03\sigma_{0},}
\end{equation}
We have assumed $\rho\sim 2.02$ to obtain the above values, and $\sigma_0$ has been determined in \eqref{crosssect}. 
To find the event horizon radius up to linear order in $\beta$, we write the following expansions for $r_{+}=r^{(0)}_{+}+\beta r^{(1)}_{+}$. Substituting this into the $f=0$, one can obtain 
\begin{equation}
    r_{+}=2m-\dfrac{5\beta}{4m^5}+\mathcal{O}(\beta^2).
\end{equation}
Considering the above expression of the event horizon, the temperature of the black hole is given as follows,
\begin{equation}
    T=\left.\dfrac{1}{4\pi}\dfrac{(Nf)^{\prime}}{\sqrt{N}}\right\vert_{r_{+}}=\dfrac{1}{8\pi m}+\dfrac{\beta}{16\pi m^{7}}+\mathcal{O}(\beta^2),
\end{equation}
where $^\prime$ denotes differentiation with respect to $r$. Finally, using this definition for the temperature, the mass loss rate reads
\begin{equation}
\dfrac{dm}{dt}\sim \dfrac{27}{4096}\dfrac{\rho \xi}{\pi^3 m^2}+\dfrac{18851}{1492992}\dfrac{\rho \xi \beta}{\pi^3 m^{8}}+\mathcal{O}(\beta^2)\,.
\end{equation}
here $\xi=\pi^2/60$.
By integrating the above expression one, obtains the lifetime of the black hole as,
\begin{align}
t=&\dfrac{4096}{81}\dfrac{\pi^3 m^3}{\xi \rho}-\dfrac{2048\pi^4\sqrt{113106\beta}}{19683\xi\rho}+\dfrac{154427392\pi^3\beta}{1594323\xi\rho m^3}\nonumber\\
&+c_{1}+\mathcal{O}(\beta^2).
\end{align}
Let us consider a black hole of initial mass $m_{0}$ created at time $t=0$ which evaporates. Using the initial condition, we can determine $c_{1}$. The primordial black holes with masses of order $m_{0}<10^{11}kg$ created in the early universe (let's say $t \approx 10Gy$) would be completely evaporated at present. Therefore, we have
\begin{equation}
    m^{3}(t)=m_{0}^{3}+\dfrac{81\xi\rho t}{4096\pi^3}+\dfrac{37702\xi\rho\beta t}{243 m_{0}^3(4096\pi^3 m_{0}^{3}+81 \xi\rho t)}.
\end{equation}
\begin{figure*}[htbp]
	\centering
	\includegraphics[width=0.95\columnwidth]{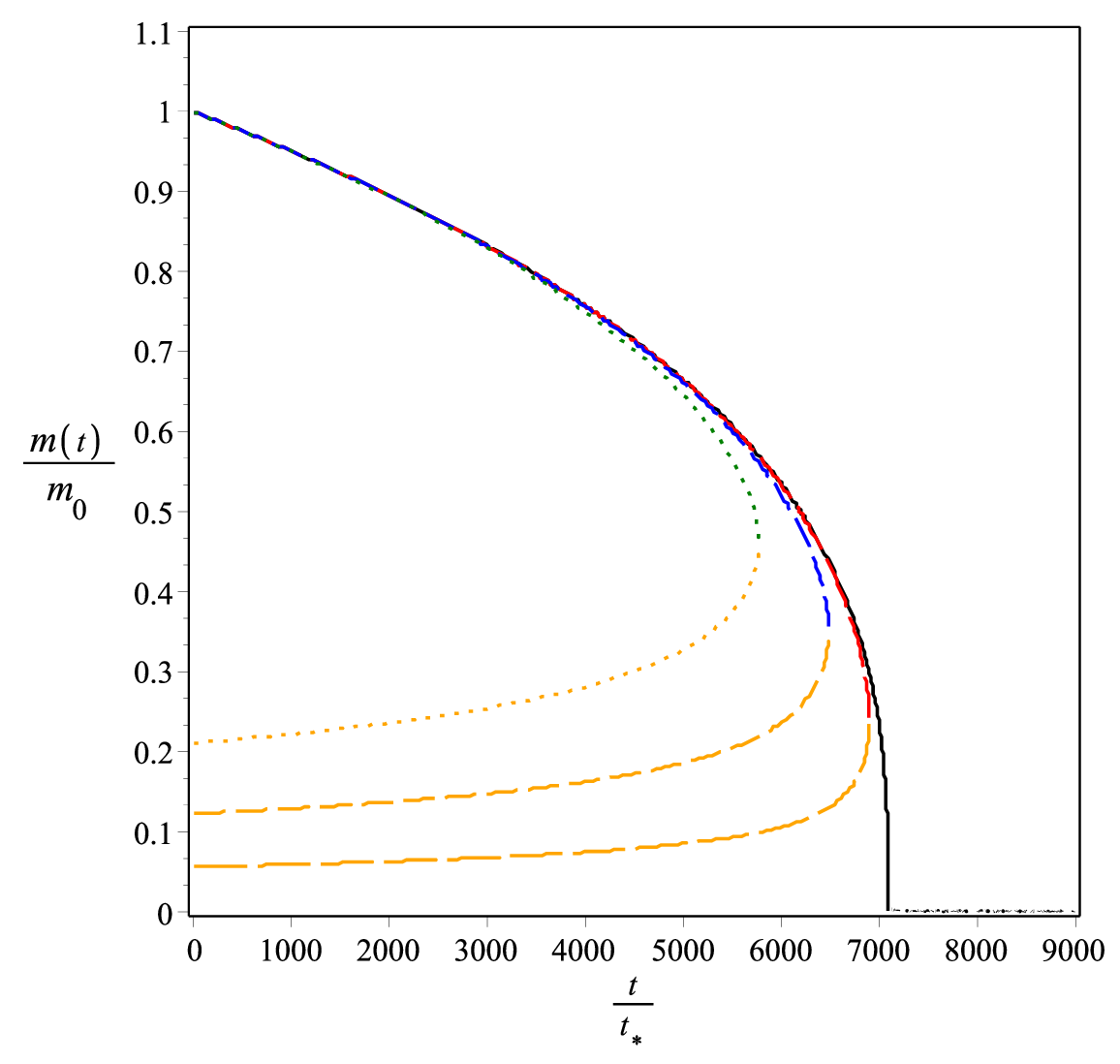}~~~~~~~~~~
	\includegraphics[width=0.95\columnwidth]{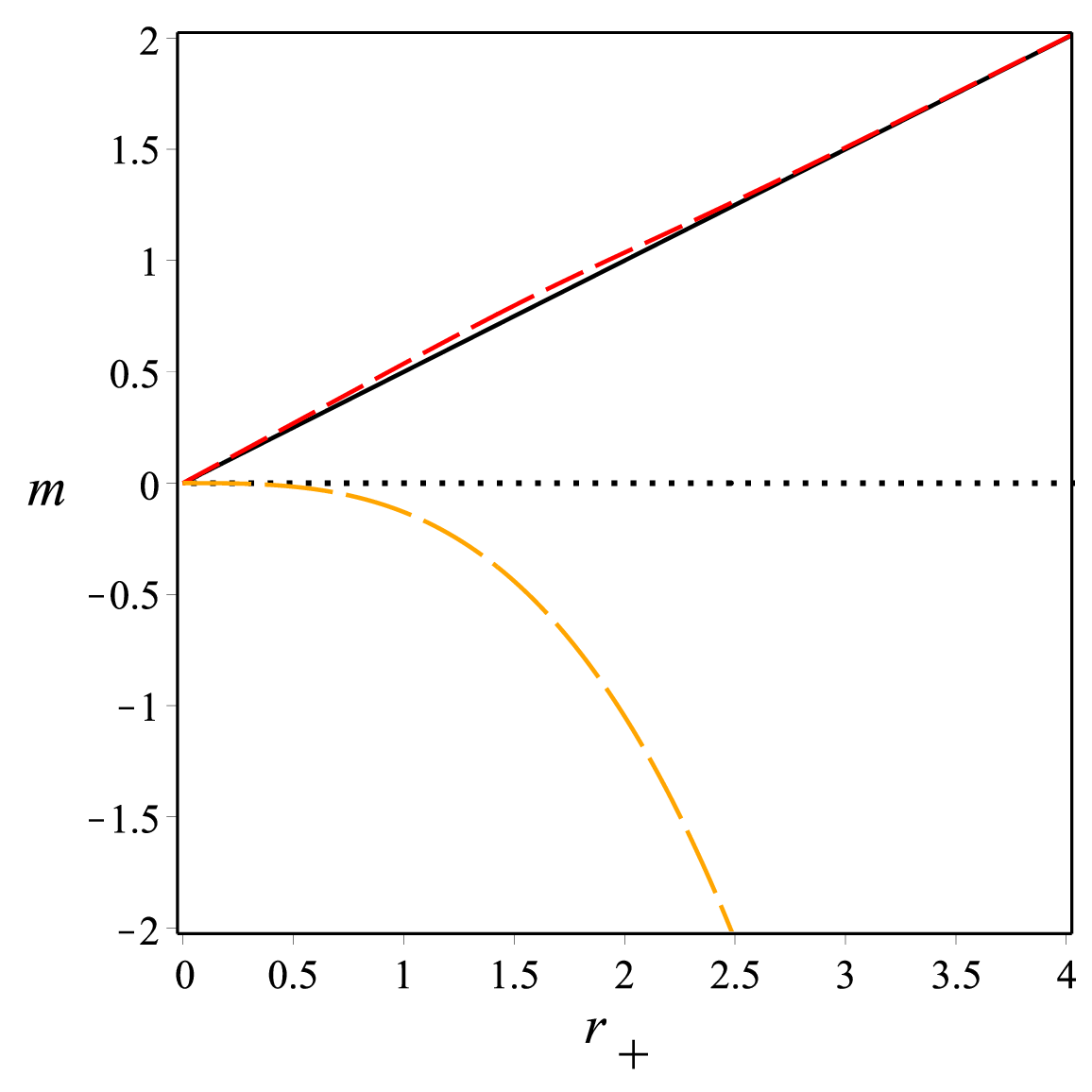}
	\caption{Evaporation of black hole with $\beta=0$ (solid line), $\beta=0.0001m_{0}^{6}$ (long dashed line), $\beta=0.001m_{0}^{6}$ (dashed line) and $\beta=0.005$ (dotted line).}
	\label{Mtevap}
\end{figure*}
Figure \ref{Mtevap} shows BH's evaporation process with different coupling theory values. We show that the mass $m(t)$ and time in units of the initial mass $m_{0}$ and $t_{\star}=m_{0}^{3}/(\xi\rho)$. As can be seen, complete evaporation occurs only for $\beta=0$, i.e. Einstein gravity. For non-zero coupling constant $\beta\neq 0$, there is no time where the mass becomes zero.

\section{Greybody factor}\label{grey}

Generally, black hole emits radiation in a form of Hawking radiation which can be treated as black-body spectrum. However, spacetime curvature outside black hole's event horizon acts as a potential barrier which leading to a deviation from black-body spectrum. Similar to a situation in quantum mechanics, the radiation can be either transmitted and reflected by the potential. Greybody factor is defined as $\gamma(\omega) \equiv T^2$, where $T(\omega)$ is transmission coefficient. This factor measures how much the radiation deviates from the black-body spectrum. The transmitted and reflected waves are defined by the following boundary conditions
\begin{align}
    \psi &\sim  \left\{
     \begin{aligned}
       & Te^{-i\omega r_*},\hspace{3cm} r_{\ast}\rightarrow -\infty \\
       & e^{-i\omega r_*} + R e^{i\omega r_*},~~~~~~~~~~~~~~~r_{\ast}\rightarrow +\infty
     \end{aligned}
   \right. 
\end{align}
and $R$ is reflection coefficient. In addition, $R(\omega)$ and $T(\omega)$ are related by $|R(\omega)|^2 + |T(\omega)|^2=1$.

Here in this section, we compute greybody factor for scalar perturbation around black hole solution in EBR gravity. We first explore the greybody factor using the rigorous bounds \cite{Visser:1998ke,Boonserm:2009zba}. The rigorous bounds offer an analytical way (or semi-analytical) to compute a lower bound on greybody factor. To do this, let us recall the following formula 
\begin{align}
    \gamma(\omega) \geq \sech^2 \left(\frac{1}{2\omega}\int_{-\infty}^{\infty} V dr_{\ast}\right),
\end{align}
Thus, for scalar and electromagnetic perturbations, we obtain the following rigorous bounds
\begin{widetext}
\begin{align}
    \gamma_{sc} &\geq \gamma_{sc_b} \equiv \sech^2\left[\frac{1}{2\omega}\left( \frac{l(l+1)}{r_+} + \frac{m}{r_+^2} + \left(\frac{499290m^4}{11r_+^{11}} -\frac{128\left(7l(l+1)+99\right)m^3}{5r_+^{10}}\right)\beta\right)\right] + \mathcal{O}(\beta^2), \nonumber \\
    \gamma_{e} &\geq \gamma_{e_b} \equiv \sech^2\left[\frac{l(l+1)}{2\omega}\left(\frac{896m^3\beta}{5r_+^{10}} - \frac{1}{r_+}\right)\right] + \mathcal{O}(\beta^2),
\end{align}
\end{widetext}
respectively.  In addition to rigorous bounds, greybody factor can also be computed via WKB method. The transmission coefficient can be approximately given by 
\begin{align}
    T(\omega) &= \left(1 + e^{2i\pi\alpha}\right)^{-1/2}, \label{transmissioncoeff}
\end{align}
where $\alpha$ is obtained from the WKB formula
\begin{align}
    \alpha &= i\frac{\left(\omega^2-V_0\right)}{\sqrt{-2V''_0}} + \sum_{i=2}^{6}\Lambda_i,
\end{align}
where $V_0$ is the effective potential evaluated at its peak. $V''_0$ is second derivative with respect to the tortoise coordinate and is determined at the maximum point. The higher order WKB ($\Lambda_{2-6}$) can be found in \cite{Schutz:1985km,PhysRevD.35.3621,konoplya2003quasinormal}. The transmission coefficients of various black holes have been investigated using the WKB method \cite{Chakrabarty:2018skk,Konoplya:2020cbv,Boonserm:2021owk,Fu:2022cul,Ma:2024pyc}.
\begin{figure*}[htbp]
\centering
\includegraphics[width=0.95\columnwidth]{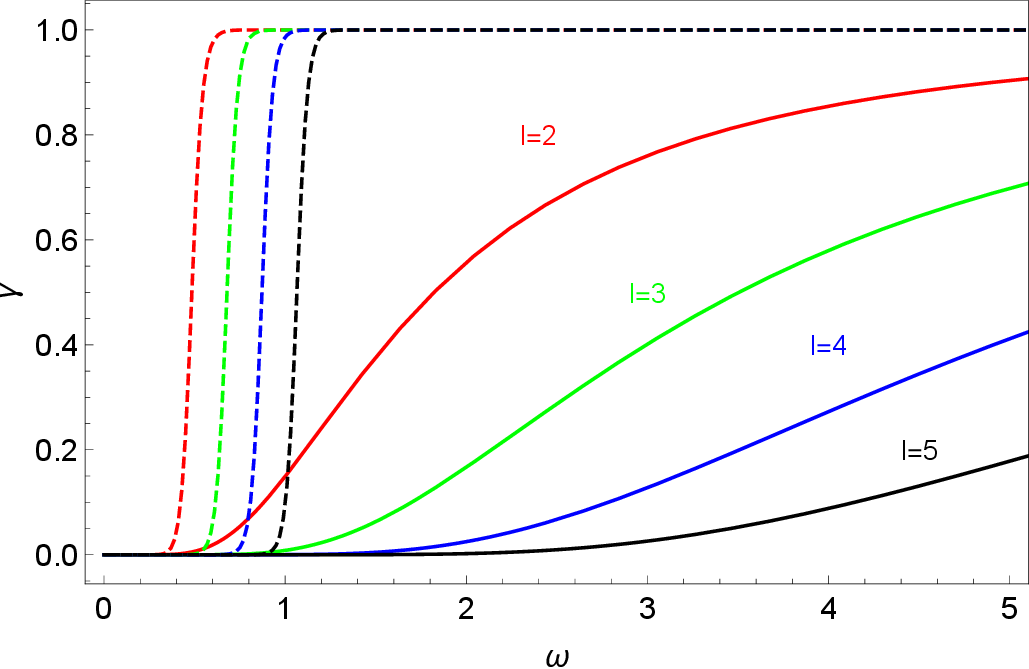}~~~~~~~~~~
\includegraphics[width=0.95\columnwidth]{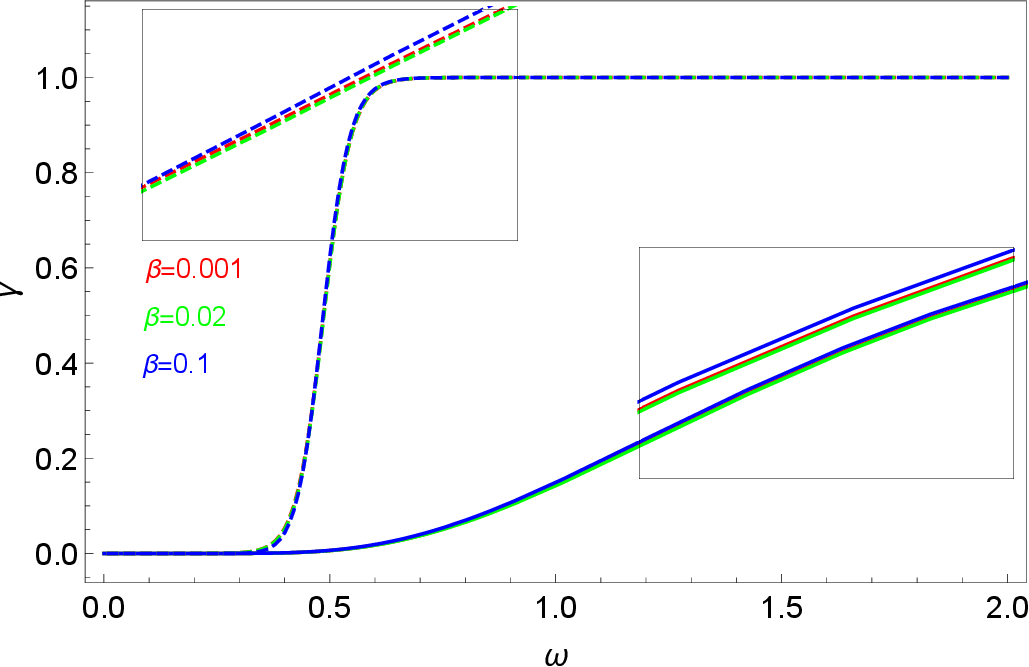}
\caption{A comparison between greybody factor for scalar perturbation obtained by rigorous bounds $\gamma_{sc_b}$ (solid lines) and WKB method (dashed lines). Left: $m=1,\beta=0.1$ Right: $m=1,l=2$ subplots display a closer behaviour of greybody factor.}
	\label{fig:rigboundSC}
\end{figure*}

\begin{figure*}[htbp]
\centering
\includegraphics[width=0.95\columnwidth]{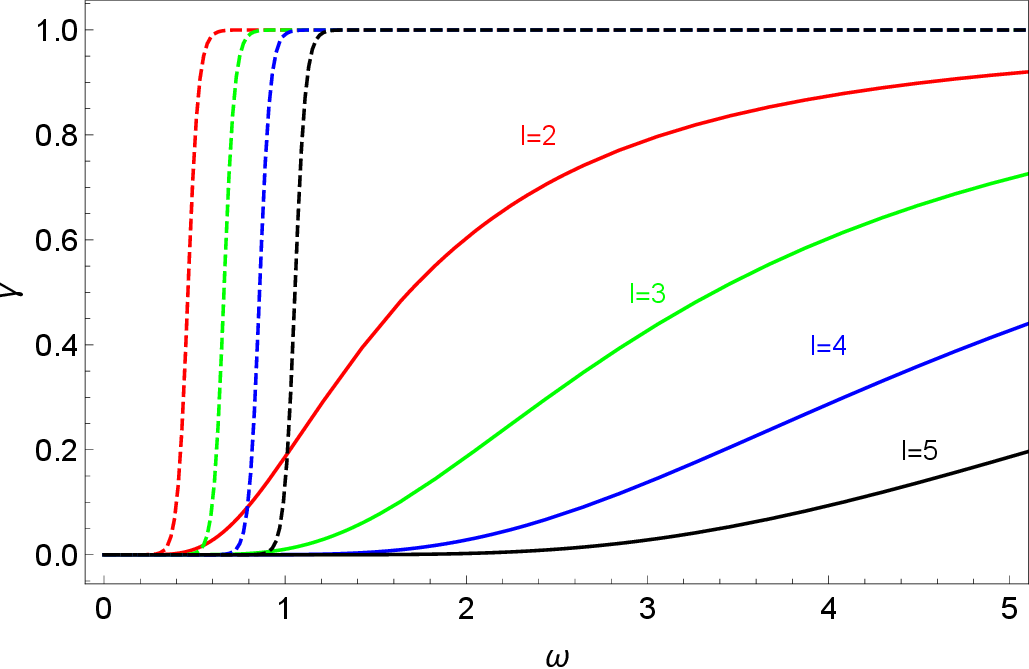}~~~~~~~~~~
\includegraphics[width=0.95\columnwidth]{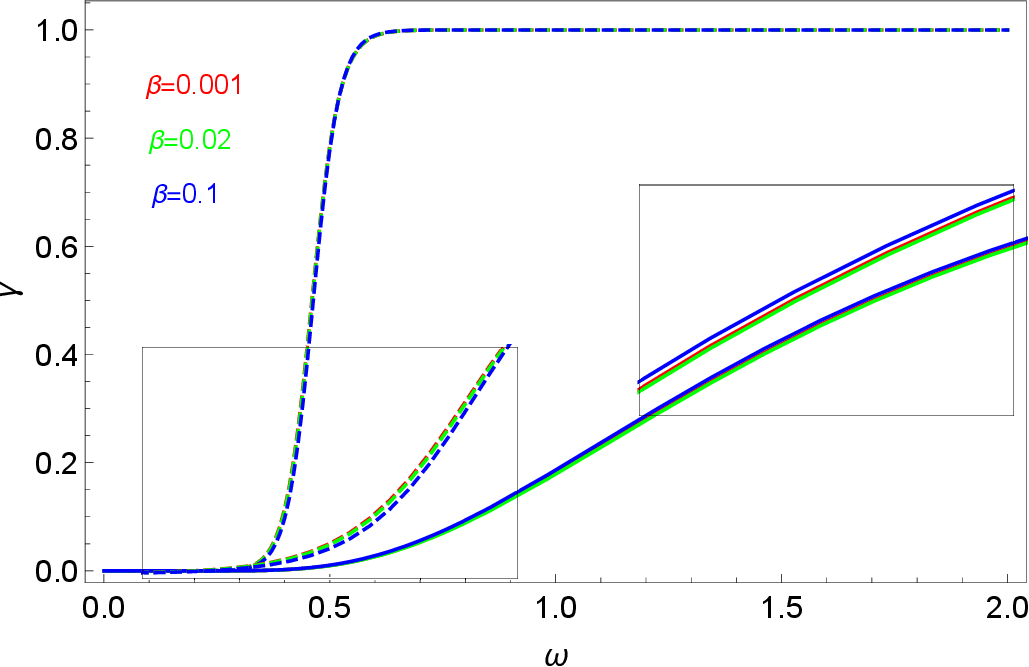}
\caption{A comparison between greybody factor for electromagnetic perturbation obtained by rigorous bounds $\gamma_{e_b}$ (solid lines) and WKB method (dashed lines). Left: $m=1,\beta=0.1$ Right: $m=1,l=2$ subplots display a closer behaviour of greybody factor. }
	\label{fig:rigboundEM}
\end{figure*}

In Fig.~\ref{fig:rigboundSC}-\ref{fig:rigboundEM}, we demonstrate the behaviour of the greybody factor for scalar and electromagnetic perturbation as a function of $\omega$, respectively. In these figures, the greybody factor is obtained via the rigorous bounds i.e., $\gamma_{sc_b}\gamma_{e_b}$ (solid lines) and the WKB method (dashed lines). In general, greybody factors for scalar and electromagnetic perturbations share similar features. 
It is clear that the rigorous bounds provide a good lower bound for the greybody factor. By increasing $l$, the greybody factor shifts toward a larger frequency $\omega$. This means that the particle requires more energy to transmit through the effective potential. This is because the higher the $l$, the higher the peak of the effective potential. In addition, increasing $\beta$ also renders a similar result as can be seen in the right panel of these figures. However, the effect of $\beta$ on the greybody factor is minuscule. Subfigures reveal the effect of $\beta$ on $\gamma$ at a close distance. Lastly, we find that greybody factor for scalar perturbation is generally larger than those of electromagnetic perturbation at a given $\omega$. This difference, however, is smaller as $l$ increases. 

Recently, a correspondence between quasinormal modes and greybody factor is reported \cite{Konoplya:2024lir}. The correspondence becomes exact in the eikonal limit and solely relies on fundamental quasinormal modes ($n=0$). For smaller $l$, correction terms are required with overtone frequency mode ($n=1$) \cite{Konoplya:2024lir}. Later, this correspondence is extended for axially symmetric spacetime \cite{Konoplya:2024vuj}. By utilising this relation, greybody factors for electromagnetic and gravitational perturbation of dilaton black hole are computed in \cite{Dubinsky:2024vbn}. 

Here, we explore the relationship between greybody factor and quasinormal modes for scalar and electromagnetic perturbation of static BH in EBR gravity. The corresponding is given by \cite{Konoplya:2024lir}
\begin{widetext}
\begin{align}
    \alpha &= -i\left(\frac{\omega^2-Re(\omega_0)^2}{4Re(\omega_0)Im(\omega_0)}\right) + i \left(\frac{Re(\omega_0)-Re(\omega_1)}{16 Im(\omega_0)}\right) -i \left( \frac{\omega^2-Re(\omega_0)^2}{32Re(\omega_0)Im(\omega_0)}\right)\left[\frac{\left(Re(\omega_0)-Re(\omega_1)\right)^2}{4Im(\omega_0)^2} \right. \nonumber \\
    &~~~~\left. - \frac{3Im(\omega_0)-Im(\omega_1)}{3Im(\omega_0)}\right] + i \frac{\left(\omega^2-Re(\omega_0)^2\right)^2}{16Re(\omega_0)^3Im(\omega_0)}\left[1+\frac{Re(\omega_0)\left(Re(\omega_0)-Re(\omega_1)\right)}{4Im(\omega_0)^2}\right] \nonumber \\
    &~~~~-i\frac{\left(\omega^2-Re(\omega_0)^2\right)^3}{32Re(\omega_0)^5Im(\omega_0)}\left[1 + \frac{Re(\omega_0)\left(Re(\omega_0)-Re(\omega_1)\right)}{4Im(\omega_0)^2} + \right. \nonumber \\
    &~~~~\left. + Re(\omega_0)^2\left(  \frac{\left(Re(\omega_0)-Re(\omega_1)\right)^2}{16Im(\omega_0)^4} - \frac{3Im(\omega_0)-Im(\omega_1)}{12Im(\omega_0)}\right)\right] + \mathcal{O}\left(l^{-3}\right), \label{greyQNMs}
\end{align}
\end{widetext}
where $\omega_0$ and $\omega_1$ are fundamental and first overtone quasinormal modes, respectively. Remarkably, one can compute transmission coefficient (hence greybody factor) by plugging the above equation into \eqref{transmissioncoeff}. 

\begin{figure*}[htbp]
\centering
\includegraphics[width=0.95\columnwidth]{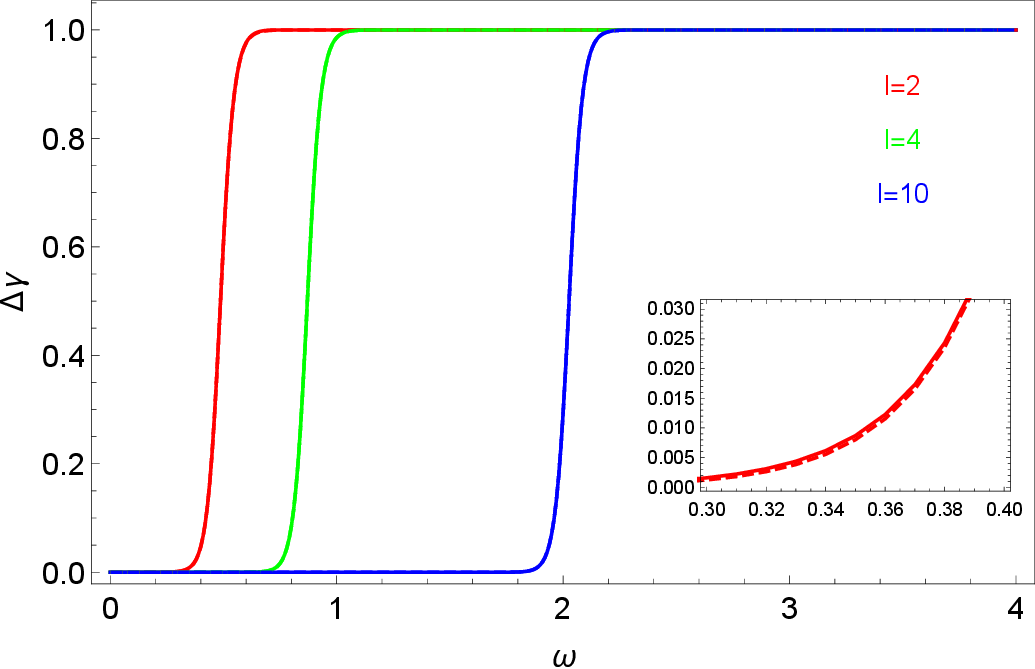}~~~~~~~~~~
\includegraphics[width=0.95\columnwidth]{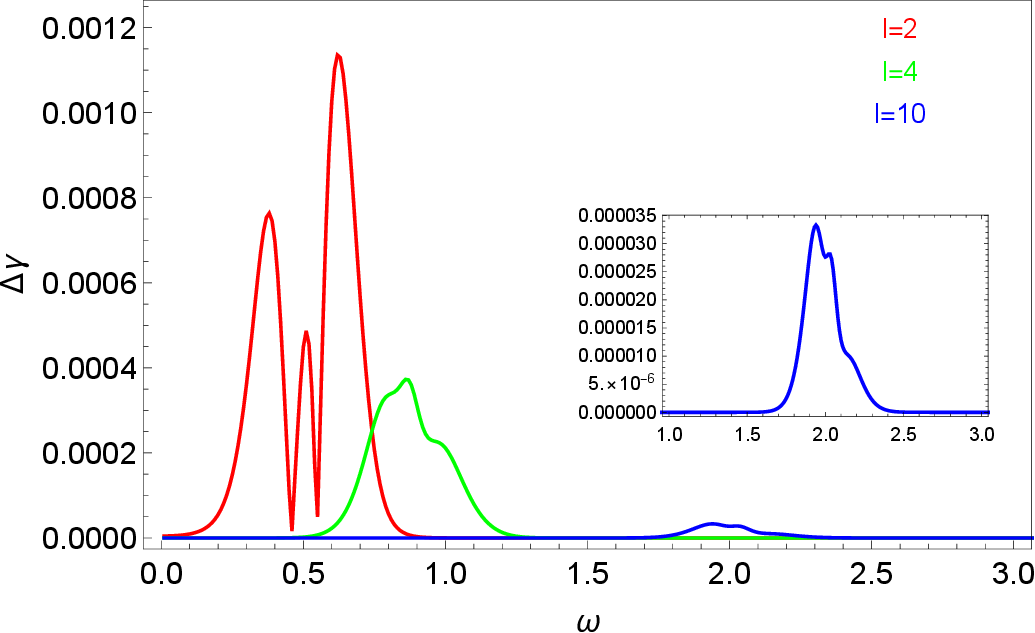}
\caption{Greybody factor for scalar perturbation as function of $\omega$ for $m=1$ and $\beta=0.05$. Left: Greybody factor computed via 6th order WKB (solid curves) and \eqref{greyQNMs} (dashed curves). Subplot shows a closer look for $l=2$ in the intermediate region. Right: The absolute difference between two methods as a function for $\omega$. Subplot shows a closer look for $l=10$ case.}
	\label{fig:delgreySC}
\end{figure*}

\begin{figure*}[htbp]
\centering
\includegraphics[width=0.95\columnwidth]{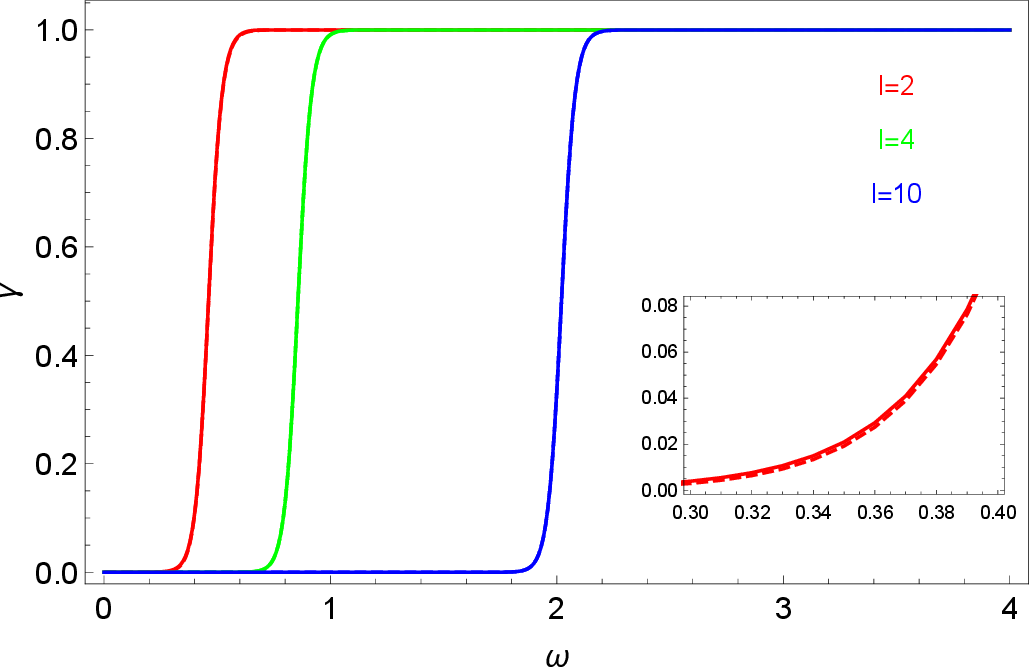}~~~~~~~~~~
\includegraphics[width=0.95\columnwidth]{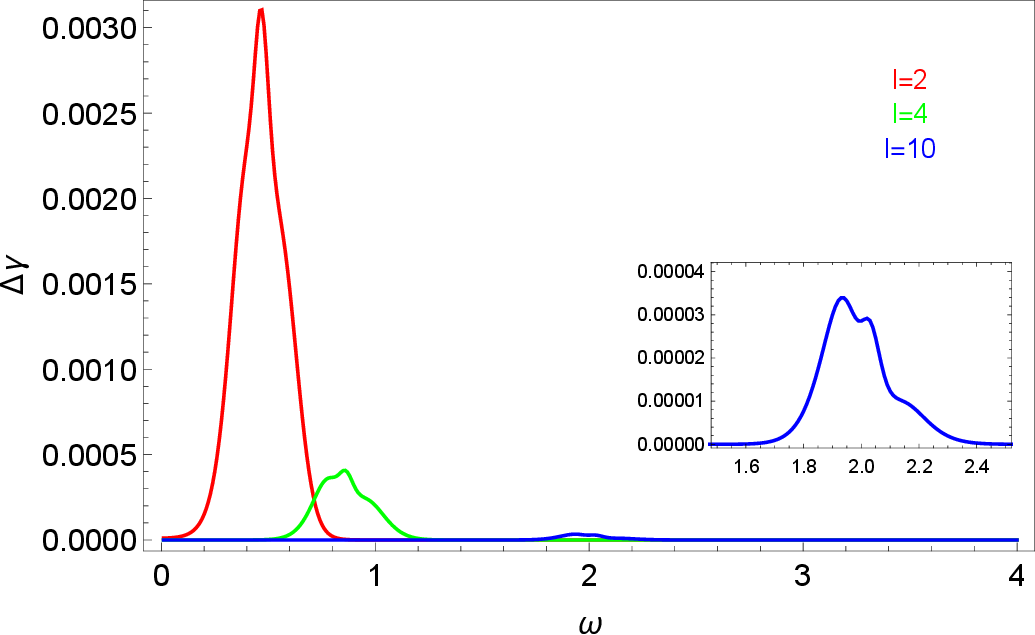}
\caption{Greybody factor for electromagnetic perturbation as function of $\omega$ for $m=1$ and $\beta=0.05$. Left: Greybody factor computed via 6th order WKB (solid curves) and \eqref{greyQNMs} (dashed curves). Subplot shows a closer look for $l=2$ in the intermediate region. Right: The absolute difference between two methods as a function for $\omega$. Subplot shows a closer look for $l=10$ case.}
	\label{fig:delgrey}
\end{figure*}

We show the difference of greybody factor calculated via 6th order WKB method and formula above \eqref{greyQNMs} in Fig.~\ref{fig:delgreySC}-\ref{fig:delgrey} for scalar and electromagnetic perturbation, respectively. In the left panel of these figures, we demonstrate the greybody factor as function of $\omega$ computed by the two previously mentioned methods. The difference merely noticeable at least for $l=2$ case (see the subfigure). In addition, we show the absolute difference i.e., $\Delta\gamma$ in the right panel of these figures. The maximum discrepancies are less than 0.0012 (scalar) and 0.0032 (electromagnetic) for $l=2$ and increasingly smaller for $l=4,10$. It is clear that $\Delta\gamma$ becomes smaller as $l$ increases. Comparing between scalar and electromagnetic perturbation, the $\Delta\gamma$ of the former case is overall lower than the latter case.


\section{Conclusion}\label{concl}
This paper studies the physical properties of static spherically symmetric EBR gravity to explore how the BR correction affects the solution.
Working to leading order in $\beta$, we find that ISCO for a massive test body will be on a smaller radius for a larger value of the coupling constant $\beta$ of EBR. Also, the angular momentum, energy of a particle, and proper period of the body at the ISCO decrease as $\beta$ increases. Likewise, the study of light-like geodesics reveals that EBR reduces the photon sphere radius, the shadow of the black hole. Light deflection near the EBR black hole decreases by increasing the EBR coupling constant. We obtain a constrained value on the EBR coupling constant using the Shapiro time delay. 
Then, using the mass loss rate of Hawking radiation we obtained the lifetime of the black hole. 
Finally, we obtain the real and imaginary parts of the quasi-normal mode in the Eikonal limit. We find that increasing the theory's coupling constant increases the real part but decreases the imaginary part. We also show that the real part of QNM is inversely proportional to the shadow radius. Moreover, to obtain a better idea of how the coupling constant impacts the QNMs spectra, we considered Pad\'e averaged sixth order WKB approximation method. The results show that the presence of the coupling constant decreases the damping rate of ring-down GWs. In the case of scalar QNMs, an increase in the coupling constant results in a decrease of the oscillation frequency of GWs while in the case of electromagnetic QNMs, the oscillation frequency increases with an increase in the coupling constant of the theory. These results highlight the crucial role of the coupling constant in modifying both the damping and frequency characteristics of QNMs. By suppressing damping and inducing distinct frequency shifts depending on the perturbation type, the coupling constant provides deeper insights into black hole stability and dynamics within the theoretical framework under consideration. The rigorous bound of greybody factors is calculated. We find that they provide a good lower bound when comparing to the greybody factors computed via 6th-order WKB. In addition, we learn that the coupling constant barely affects the greybody factor. Lastly, the corresponding between greybody factor and quasinormal modes is confirmed (up to first order in $\beta$).

A natural direction for future work would involve extending these results to compute shadows of rotating black holes in EBR. These are of more direct astrophysical relevance and may present distinct angular-dependent features that could be observed. Similar techniques as those presented here could be used to obtain approximate rotating black hole solutions in this theory.

\section*{Acknowledgements}
This research has received funding support from the NSRF via the Program Management Unit for Human Resource and Institutional Development, Research and Innovation grant number $B13F670063$. 

\appendix
\section{Explicit Terms in the Continued Fraction Approximation}\label{sec:Appendix}

We present terms up to {fourth order} in the continued fraction approximation \eqref{cfrac}: 
\begin{align}
&\epsilon=-\dfrac{F_1}{r_+}-1,\,\,\,\, a_1=-1-a_{0}+2\epsilon+r_{+}h_1,\nonumber\\
&a_{2}=-\dfrac{1}{ a_1} \left[4a_1-5\epsilon+1+3 a_{0}+ h_{2}r_+^2\right]
\nonumber \\
&a_{3}=-\dfrac{1}{{a_1}{a_{2}}}\Big[-{h_{3}}r_+^3+{a_1}a_2^2+5{a_1}{a_{2}}+6{a_{0}}+10{a_1}-9\epsilon+1\Big],\nonumber\\
&a_{4}=-\dfrac{1}{a_1a_{2}a_{3}}
\Bigg[h_{4}r_{+}^2+a_1a_{2}^3+2a_1a_{3}a_{2}^2+a_1a_{2}a_{3}^2+6a_1a_{2}^2+\nonumber\\
    &~~~~~~~~~~6a_1a_{2}a_{3}+15a_1a_{2}+10a_{0}+20a_1-14\epsilon +1\Bigg],
\end{align}
and 
\begin{align} 
b_1 &= -1+\sqrt{\dfrac{h_1}{f_1}},\nonumber\\
b_{2}&=\dfrac{(-4f_{1}+f_{2}r_{+})b_{1}^2+2(-2f_{1}+f_{2}r_{+})b_{1}+r_{+}(f_{2}-h_{2})}{2f_{1}b_{1}(1+b_{1})},\,\,\nonumber\\
b_{3} &= \dfrac{1}{2f_{1}b_{1}b_{2}(1+b_{1})}\Big[(-f_{3}r_{+}^2+2f_{2}r_{+}(2+b_{2})-f_{1}(10+3b_{2}^2+\nonumber\\
&10b_{2}))b_{1}^2+(-2f_{3}r_{+}^2+2f_{2}r_{+}(2+b_{2}-2f_{1}(3+b_{2}^2+3b_{2})))b_{1}\nonumber\\
&-r_{+}^2(f_{3}-h_{3})\Big]\nonumber\\
b_{4} &=\dfrac{1}{2f_{1}b_{1}b_{2}b_{3}}\Big[(-4f_{1}b_{2}^3+b_{2}^{2}(3f_{2}r_{+}-6f_{1}(b_{3}+3))+{b_{2}}(-2f_{3}r_{+}^{2}\nonumber\\
&+2f_{2}r_{+}(b_{3}+5)-2f_{1}(6b_{3}+b_{3}^2+15))+f_{4}r_{+}^3-20f_{1}-4f_{3}r_{+}^2\nonumber\\
&+10f_{2}r_{+}{)}b_{1}^2+(-2f_{1}b_{2}^3+(2f_{2}r_{+}-4f_{1}(b_{3}+2))b_{2}^2+(-2f_{3}r_{+}^2+\nonumber\\
&2f_{2}r_{+}(b_{3}+3)-2f_{1}(6+4b_{3}+b_{3}^2))b_{2}-8f_{1}+6f_{2}r_{+}-4f_{3}r_{+}^2+\nonumber\\
&2f_{4}r_{+}^3)b_{1}+r_{+}^{3}(f_{4}-h_{4})\Big].  \nonumber
\end{align}
\bibliography{sample}



\end{document}